\documentclass[journal]{IEEEtran}
\IEEEoverridecommandlockouts
\usepackage[caption=false]{subfig}
\usepackage{url}
\usepackage{graphicx}
\usepackage{cite}
\usepackage[english]{babel}
\usepackage{amsmath,amssymb,amsfonts}
\usepackage{tikz}
\usepackage[linesnumbered,ruled,noline]{algorithm2e}%{algorithmic}
\usepackage{bm}
\usepackage{bbm}
\usepackage{tabularx}
\usepackage[normalem]{ulem}
\graphicspath{{./figures/}}
\newboolean{trackchanges}

\renewcommand{\vec}[1]{\bm{#1}}
\DeclareMathOperator{\E}{\mathrm{E}}
\newcommand{\cond}{\,|\,}

\begin{document}
\title{Policy Gradient Algorithms for Age-of-Information Cost Minimization}

\author{Jos\'e-Ram\'on~Vidal, Vicent~Pla, Luis~Guijarro, and Israel~Leyva-Mayorga,~\IEEEmembership{Member, IEEE}~\thanks{Jos\'e-Ram\'on~Vidal, Vicent~Pla and Luis~Guijarro are with the Department of Communications, Universitat Polit\`ecnica de Val\`encia, Cam\'i de Vera s/n, 46022 Val\`encia, Spain (e-mail: jrvidal@upv.es; vpla@upv.es; lguijar@dcom.upv.es)}
~\thanks{Israel Leyva-Mayorga is with the Department of Electronic Systems, Aalborg University, 9220 Aalborg, Denmark (e-mail: ilm@es.aau.dk)}\thanks{\copyright 2025 IEEE. Personal use of this material is permitted. Permission from IEEE must be obtained for all other uses, in any current or future media, including reprinting/republishing this material for advertising or promotional purposes, creating new collective works, for resale or redistribution to servers or lists, or reuse of any copyrighted component of this work in other works.}}

\markboth{IEEE Transactions on Communications}{Vidal \MakeLowercase{\textit{(et al.)}}: Policy Gradient Algorithms for Age-of-Information Cost Minimization}

\maketitle
\begin{abstract}
Recent developments in cyber-physical systems have increased the importance of maximizing the freshness of the information about the physical environment. However, optimizing the access policies of Internet of Things %(IoT)
devices to maximize the data freshness, measured as a function of the Age-of-Information (AoI) metric, is a challenging task. 
This work introduces two algorithms to optimize the information update process in cyber-physical systems operating under the \textit{generate-at-will model},
 by finding an online policy without knowing the characteristics of the transmission delay or the age cost function.
The optimization seeks to minimize the time-average cost, which integrates the AoI at the receiver and the data transmission cost, making the approach suitable for a broad range of scenarios. Both algorithms employ policy gradient methods within the framework of model-free reinforcement learning (RL) and are specifically designed to handle continuous state and action spaces. Each algorithm minimizes the cost using a distinct strategy for deciding when to send an information update. Moreover, we demonstrate that it is feasible to apply the two strategies simultaneously, leading to an additional reduction in cost.
The results demonstrate that the proposed algorithms exhibit good convergence properties and achieve a time-average cost within 3\% of the optimal value, when the latter is computable. 
A comparison with other state-of-the-art methods shows that the proposed algorithms outperform them in one or more of the following aspects: being applicable to a broader range of scenarios, achieving a lower time-average cost, and requiring a computational cost at least one order of magnitude lower.
\end{abstract}


%--------------------------------------------------------------
%--------------------------------------------------------------
\section{Introduction}\label{introduction}
%--------------------------------------------------------------
%--------------------------------------------------------------

\IEEEPARstart{N}{umerous} Internet of Things (IoT) applications such as remote surgery and health monitoring, intelligent transportation systems, intelligent energy  grid management and others are realized through cyber-physical systems, which 
enable seamless interaction between a virtual domain and the surrounding physical environment 
and require networked interaction among computing and physical components.
An IoT network supporting these applications typically consists of multiple wireless sensors 
which measure underlying physical stochastic processes and transmit updates of the obtained measurements to destination nodes 
that keep track of the status of these processes.
In the nodes, the information received is processed by controllers that generate control
commands and then transmit them to the actuators.
When communication between sensors, actuators, and controllers is performed wirelessly, 
we refer to these applications as wireless networked control systems.
Because of the transient nature of data, sensor measurements become outdated over time. 
The performance of wireless networked control systems depends upon how fresh the status updates are when they reach the destination nodes. 
In addition, the delivery of status updates may be restricted 
by the limited energy supply of the source nodes and the transmission delay of the communication network between the source and destination nodes. 
The delay and loss of updates could result in staleness of information status at the destination nodes, 
which eventually degrades the performance of such real-time applications.
This calls for update techniques in which sensors sample and send
a minimal number of measurements to conserve the energy
while still providing the end users with highly fresh data, as required by time-sensitive applications~\cite{kah23}. 

The freshness of information can be quantified by  the age of information (AoI) metric introduced in~\cite{kaul12}. 
The AoI  is a performance measure to characterize the data staleness at the  destination node,
and it is defined as the time elapsed since the most recently received status update 
at the destination node  was generated at the source node. 
The AoI does not rely on the usual performance measures, such as the transmission delay or throughput,
but it rather measures the  delay in updating the status of the underlying physical process at the destination node.

In recent years, AoI minimization has been studied in various scenarios and under different assumptions~\cite{yates21}. 
In those scenarios in which the ability of sensors to make updates is limited by external factors,
such as energy consumption, the \textit{generate-at-will} model is usually adopted.
In this model it is assumed that  the source can generate and transmit a status update at any time (i.e., on demand). 
This model has also been  explored in the context of sampling, estimation and control.
With generate-at-will sources, the long-term time-average of the AoI can be minimized by choosing the optimal instants to generate new updates.
In general, to minimize the time-average AoI, 
the source node often has to wait before generating and sending a new update even when the channel is currently idle.
The optimal policy to choose the update times depends on the transmission characteristics, such as losses and delay distribution.
It also depends on how the AoI measure is defined,
since the AoI measure to be minimized can be a generalization of the basic AoI definition~\cite{sun19}.
Besides, other constraints in the problem definition may limit the frequency of the updates,
such as energy limitations~\cite{cao23}, or any other costs associated to the transmission of updates.

In most practical systems, the transmission delay distribution is not known a priori, and must be learned.
Besides, the function that quantifies the AoI cost depends on the specific application that is currently running at the destination node, 
and may be unknown to the source node, so it will  also have to be learned.
For this reason, there is growing interest in applying learning-based methods to this problem~\cite{kah23}.
In this work, we consider a model-free approach based on RL, which decides the
updating times based on the system state to optimize a time-average cost that includes the AoI, 
as well as other factors, and learns the updating policy by interacting with the environment. 

The main contributions of this paper are the following:
\begin{enumerate}
\item We present two versions of a novel policy gradient algorithm to reduce a general time-average cost based on AoI,
with good convergence properties and low complexity. 
\item We propose two distinct update strategies to reduce the time-average cost.
\item We demonstrate the applicability of this algorithm for a wide range of cost functions and for different strategies to reduce the AoI cost.
\item We show the capability of the algorithm to operate when several update strategies are implemented simultaneously, 
without increasing noticeably the computational cost.
\end{enumerate}

This paper is structured as follows.
In Section~\ref{sec:relwork}, we review the related work in  AoI minimization under the  \textit{generate-at-will} model, 
and in the use of RL to address the problem.
Section~\ref{sec:model} formulates the model of the system.
In Section~\ref{sec:algs}, the learning setup is described and two versions of the proposed algorithm are presented. 
In Section~\ref{sec:results}, results are presented and discussed. 
Finally, Section~\ref{sec:conclusion} draws the conclusions.

\section{Related work}\label{sec:relwork}

\noindent The AoI minimization problem under the \textit{generate-at-will} model was addressed in~\cite{sun17}.
In this work, the authors show that submitting a fresh update 
immediately after the previous update is delivered and the communications network is ready 
is not always the optimal policy to minimize the long-term time-average AoI.
Instead,  in some cases it is better to wait for a certain time before submitting a new update.
They define a possibly non linear penalty function to characterize the level of dissatisfaction for data staleness,
and formulate the time-average age penalty minimization
problem as a constrained semi-Markov decision problem,
in which the state is the transmission delay of the last delivered update.
From this, the optimal waiting time as a function of the transmission delay, 
given the penalty function and the transmission delay distribution,
is calculated by means of a two-layer bisection algorithm.
In~\cite{tsai23} the same problem is addressed, but this time it is solved by 
algorithms that can be run online. 
Furthermore, knowledge of the transmission delay distribution is not required,
nor is knowledge of the AoI penalty function, 
which can be estimated using monotonic regression.

These works solve the problem only for the update strategy described above, 
which consist of waiting a certain amount of time before generating a new sample,
and only for the case where the cost is defined as the time-average cost of a certain AoI penalty function.
However, these algorithms do not extend to other update strategies or alternative AoI cost measures, such as threshold-based metrics. 
Moreover, minimizing the time-average cost becomes infeasible when transmission costs are included---an important consideration for capturing factors like energy consumption.

Most RL approaches for minimizing the time-average AoI are applicable only to slotted time settings, 
and for  a discrete and finite action space,
and are not applicable to the problem described above.
In~\cite{hatami21}, the authors propose a Q-learning algorithm to reduce the long-term time-average AoI cost 
in an IoT  sensing network with a set of data consumers, 
a wireless edge node, and a set of energy harvesting  sensors.
Each sensor independently measures time-sensitive information of a physical process, 
and can send a status update only if it has at least one unit of energy in its battery.
The actions are taken by the edge node at every time slot, and can be either to command a sensor to perform a new measurement
and send a status update or to use the previous measurement to serve the request.

In~\cite{ceran19}, the authors address the time-average AoI minimization in a setting consisting of a  source  node 
sending updates to a destination over  an error-prone communication link and an HARQ scheme.
% that changes randomly from one time slot to the next.
%In the HARQ protocol, the receiver combines information from all previous transmission attempts of the same packet in
%order to increase the success probability of decoding [5]–[7]. The exact relationship between the probability of error and the number of retransmission attempts varies depending on the channel conditions and the particular HARQ method employed [5]–[7]. In general, the probability of successful decoding increases with each transmission, but the AoI of the received packet also increases. Therefore, there is an inherent tradeoff between retransmitting previously failed status information with a lower error probability, or sending a fresh status update with higher error probability.
%
%For the simulations employing HARQ, motivated by previous research on HARQ [5]–[7], we assume that decoding error reduces exponentially with the number of retransmission, that is, g(r) = p0·λ^r for some λ ∈ (0, 1), where p0 denotes
%the error probability of the first transmission, and r is the retransmission count (set to 0 for the first transmission).

The AoI is quantified as the number of time slots elapsed since the generation of the most up-to-date packet successfully decoded at the receiver. 
The state of the system is defined by the AoI at the beginning of the time slot and the number of previous transmission attempts of the same packet.
At each time slot, three actions are possible: idle, transmit or retransmit. 
To solve this problem, an average-cost SARSA algorithm with softmax exploration is proposed. 
The same authors propose in~\cite{ceran21} the same average-cost SARSA algorithm 
to address the scheduling of status updates in a network with a source terminal
that monitors a time-varying process and sends updates to a set of users.
At the beginning of every time slot, 
the source is able to generate an update and transmit it to at most one of the users. 
The state of each of the channels changes randomly from one time slot to the next and is unknown by the sender.
The action is taken at the beginning of every time slot, 
and consists of deciding whether to generate an update and, if so, to which user to transmit it.
In~\cite{elmagid20}, a deep RL algorithm is proposed for  minimization of the average weighted sum-AoI
in a real-time monitoring system consisting of a set of source nodes,
each of them observing a physical process and sending updates  to a destination node.
The  source nodes harvest RF energy in the downlink but cannot simultaneously harvest and transmit data in the uplink. 
Channels are assumed to remain constant over a time slot but change independently from one slot to another, and their power gains are known by the sources. 
The state of a source node is characterized by its battery level, the AoI at the destination of its observed process, and its channel power gain.
The action taken at each time slot, based on the joint state of the source nodes, 
is to allocate the slot to broadcasts an RF energy signal in the downlink or to allocate the slot to a single source to transmit an update.
All of these works are limited to slotted-time scenarios, with fixed-length slots, 
and with a limited set of actions, and are not applicable to update strategies like the one described in~\cite{sun17,tsai23}.

The problem of time-average AoI minimization in a continuous time setting, 
and with a continuous action space, 
is addressed in~\cite{kam19} by means of a model-free RL approach.
A SARSA algorithm is described, which uses tile coding to learn
and approximate an action-value function. 
With this algorithm, neither knowledge of the transmission delay distribution nor knowledge of the cost function is required.
Besides, the cost function to minimize is not limited to an AoI cost based on a penalty function, 
and other strategies to minimize the time-average cost, different from `waiting before submitting', could potentially be implemented. 
The experimental results show that this algorithm reduces the time-average cost but it fails to approximate the optimal policy.

The previously discussed studies relate to the strategy of adjusting the waiting time between the moment when the previous update has been delivered and a new update is sent. The other strategy that we consider in this paper involves discarding (or preempting, as it is usually referred to in the literature) the ongoing delivery, if it is taking too long, and then sending a fresh update.

Preemptive service disciplines and buffer management techniques have been considered as effective means to improve AoI-related metrics. Of these, buffer management techniques (sometimes referred to as preemption in waiting or replacement in the buffer) are less relevant to our work, since our focus is the generate-at-will model, and there is no buffering of updates.  Therefore, in the following, we restrict our attention to related work in preemptive service disciplines.

One of the first works to address the idea of preemption in the context of AoI is~\cite{Kaul12b}. The authors study an M/M/1 queue and show that the LFCS-P (Last Come First Served - Preemptive) discipline yields lower AoI for the whole range of system loads compared to LCFS or FCFS (First Come First Served). 
Similarly, in~\cite{Bedewy17}, the authors show that if the service times are exponentially distributed, the LFCS-P discipline is AoI-optimal in a rather general sense for an arbitrary arrival process and any queue size. 
However, in~\cite{Bedewy19}, the same authors show that the optimality of the LFCS-P discipline does not hold for non-exponential service times. A gamma-distributed service time is considered in~\cite{Najm16}, and it is shown that, in general, LCFS performs better than LCFS-P, especially for high arrival rates.

The authors of~\cite{Akar20} developed a versatile queueing model allowing them to derive steady-state distributions of AoI and Peak-AoI (PAoI) for a wide variety of service-time distributions and to study the impact of service-time variability on AoI performance with and without preemption. They observed that when the variability is high, the preemptive discipline outperforms the non-preemptive one, while the preemptive discipline performs quite poorly for small variabilities, especially for larger system loads. These results help explain or confirm the apparently contradictory conclusions of previous studies. This model is extended to a multi-source environment in~\cite{Dogan21}. 
The model in~\cite{Akar21}  also includes a quite general service-time distribution in a multi-source environment. The results there confirm that the preemptive discipline is more effective when the service-time variability is high, whereas it does not perform well with distributions with reduced variability, such as deterministic or uniform.

Fiems~\cite{Fiems23} considers a two-source model with general service times and self-preemption (only packets from the same source preempt the ongoing service), and derives the Laplace-Stieltjes transform of the time-average AoI and of the PAoI. The numerical results confirm that self-preemption is beneficial when service times are exponentially distributed, whereas it is not if the variability of the service time is low.

Service disciplines in which the preemption of the packet in service is decided dynamically on a packet-by-packet basis upon the arrival of a new packet, based on the system state, are considered in~\cite{Prandel24,Qin24}. In both cases, the aim is to find the optimal preemption policy within a certain family. 
Specifically,~\cite{Prandel24} targets the optimal mean AoI or mean PAoI, and the preemption decisions are mainly based on the mean residual service time (MRST) of the packet in service and the newly arrived one, while~\cite{Qin24} targets the optimal mean AoI, and the decision is based on the lengths of the packet in service and that of the newly arrived one, and other system state variables. Furthermore, in~\cite{Qin24}, an approach based on Markov Decision Processes (MDPs) is used, assuming that the packet interarrival time and packet length distributions are known, and it is proved that the optimal policy has a threshold structure based on packet length. The authors also propose an RL method for the case in which the packet interarrival time and packet length distributions are unknown.

All the cited works on preemptive service disciplines do not follow the generate-at-will model. Instead, they assume that updates are generated according to a certain random arrival process. In~\cite{Akar23}, both models—generate-at-will and random arrivals—are considered, but preemption (in the buffer) is only considered for the random arrivals model.
Also, while these works have developed complex models and useful analyses that help to understand the impact of preemption on AoI, only~\cite{Prandel24,Qin24} adopt an optimization perspective. However, the actual packet sizes (delivery times) in~\cite{Qin24}, and the service-time distribution in~\cite{Prandel24}, are known to the decision maker, which is another key difference in comparison to our study.

The works in~\cite{Arafa19} and~\cite{Banerjee24} are arguably the most closely related to ours. Both adopt a generate-at-will model and aim to minimize the time-average AoI. Moreover, they not only analyze preemption-based strategies but also consider the waiting strategy in conjunction with preemption. Nevertheless, these studies differ from ours in several important aspects. First, they require knowledge of the service-time distribution to obtain the optimal policy, whereas our model does not need such prior information. Second, their optimization objective is the average AoI, which corresponds to using the identity function $f(x)=x$ as the age cost, while we allow for a broad class of cost functions. Third, our model accommodates correlation between consecutive service times, whereas the referenced works on preemptive disciplines assume independent service times.

Nonetheless, despite these differences, the findings reported in existing work on preemptive disciplines provide some justification for considering a preemptive policy---particularly one with a form similar to the one we propose in this paper.

%--------------------------------------------------------------
%--------------------------------------------------------------
\section{Model}\label{sec:model}
%--------------------------------------------------------------
%--------------------------------------------------------------

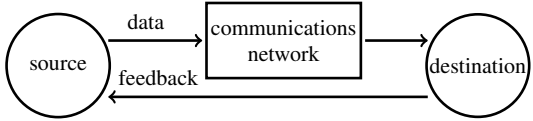
\begin{figure}[tbp]
\begin{center}
\resizebox{0.8\columnwidth}{!}{
\begin{tikzpicture}
\draw[very thick,-To](0.9,0)--(2.4,0);
\draw[very thick,-To](6,-0.9)--(0.9,-0.9);
\draw[very thick,-To](5,0)--(6,0);
\node [above,very thick,above,rectangle,inner sep=2pt,draw,anchor=center,align=center, minimum height=12mm ] at (3.7,0) {communications\\ network};
\node [very thick,above,circle,,inner sep=8pt,draw,anchor=center] at (0.1,-0.4) {source};
\node [very thick,above,circle,inner sep=1pt,draw,anchor=center] at (6.8,-0.4) {destination};
\node [anchor=center]at (1.5,0.3){data};
\node [anchor=center]at (1.7,-0.6){feedback};
\end{tikzpicture}}
\caption{Data delivery model.}\label{fig:model}
\end{center}
\end{figure}

\noindent Consider the data delivery model in Fig.~\ref{fig:model}. 
The data updates are generated according to the generate-at-will model, 
that is, at any time $t \in \mathbb{R}_{>0}$ the source can generate an up-to-date data unit and send it to the destination. 
As in~\cite{sun17,chen24}, data units are transmitted and delivered with a random transmission delay, which accounts for all communication layer functions (e.g., flow and congestion control, medium access, etc.). 
In these works, as in most of the present work, the source receives instantaneous feedback and therefore knows the transmission delay of each delivered data unit.
However, as described in Section~\ref{sec:generalization}, the proposed algorithm can be easily adapted to operate under a channel model with feedback delay (Fig.\ref{fig:modelF}), such as the one considered in\cite{tsai23}.
After receiving the feedback, the source can generate and send another data unit. 

Between two consecutive data deliveries, an AoI cost is accrued by the system due to the aging of the most recently received information. 
We assume that in cases where the AoI cost cannot be calculated at the source from the transmission delays, the source receives feedback for every delivery, containing the AoI cost as perceived by the receiver over the interval between the previous and current delivery.

Let us call {\em delivery} each time a data unit is delivered at the destination, and let $D_i$, with $i=1\dots\infty$, 
be the time of the $i$-th delivery, and set $D_0=0$.
We assume that the source can cancel the transmission of a data unit at any time before it is delivered,
so in general a delivery may require multiple data-unit acquisitions and transmissions.
 
Let $R_{(i,j)}$ be the acquisition time, or time at which the source generates the $j$-th data unit  transmitted in delivery $i$,
with $j=1\dots k_i$, 
where $k_i$ is the number of data units transmitted in delivery $i$.
Of these, only the $k_i$-th data unit is delivered.
Let $Y_{(i,j)}$ denote the transmission delay of the $j$-th data unit in the $i$-th delivery, 
measured from the time the data unit is generated at the source at  $R_{(i,j)}$ 
until the time it would be delivered to the destination.

As in~\cite{sun17}, we assume that the transmission  delays  of consecutive data units, 
	$\{Y_{(i,j)}\}_{i\geq 1, j=1,\ldots, k_i}$,
	form a stationary and ergodic Markov chain with a possibly uncountable state space, 
	and that the delays have a positive and finite mean, $0 < \E[Y_{(i,j)}] < \infty$\label{MC_assumption}.
From $D_{i-1}$ to $D_i$, only the $k_i$-th data unit is delivered at time $D_i$.
If no data unit transmission is canceled, $k_i=1$.
To simplify the notation, let $R_i$ and $Y_i$ represent the acquisition and transmission times, respectively, 
of the data unit that is actually delivered in delivery $i$.
Specifically, $R_i=R_{(i,k_i)}$, $Y_i=Y_{(i,k_i)}$, and $Y_i=D_i-R_i$.

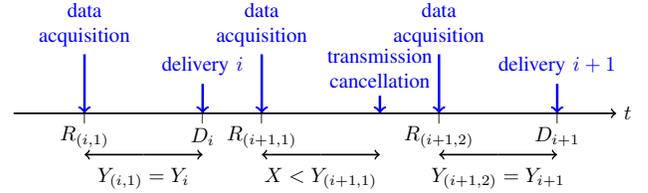
\begin{figure}[tbp]
\begin{center}
\resizebox{0.95\columnwidth}{!}{
\begin{tikzpicture}
\node [anchor=center]at (1,0){$|$};
\node [anchor=center]at (1,-0.4){$R_{(i,1)}$};
\draw[very thick,<-,color=blue] (1,0) -- (1,1) node[above, align=center] {data \\ acquisition};
\node [anchor=center]at (3,0){$|$};
\node [anchor=center]at (3,-0.4){$D_i$};
\draw[very thick,<-,color=blue] (3,0) -- (3,0.5) node[above, align=center] {delivery $i$};
\draw[thick,-To](2,-0.7)--(1,-0.7);
\draw[thick,-To](2,-0.7)--(3,-0.7);
\node [anchor=center]at (2,-1.1){$Y_{(i,1)}=Y_i$};
\node [anchor=center]at (4,0){$|$};
\node [anchor=center]at (4,-0.4){$R_{(i+1,1)}$};
\draw[very thick,<-,color=blue] (4,0) -- (4,1) node[above, align=center] {data \\ acquisition};
\node [anchor=center]at (7,0){$|$};
\node [anchor=center]at (7,-0.4){$R_{(i+1,2)}$};
\draw[thick,-To](5,-0.7)--(4,-0.7);
\draw[thick,-To](5,-0.7)--(6,-0.7);
\node [anchor=center]at (5,-1.1){$X<Y_{(i+1,1)}$};
\draw[very thick,<-,color=blue] (6,0) -- (6,0.3) node[above, align=center] {transmission\\ cancellation};
\draw[very thick,<-,color=blue] (7,0) -- (7,1) node[above, align=center] {data \\ acquisition};
\node [anchor=center]at (9,0){$|$};
\node [anchor=center]at (9,-0.4){$D_{i+1}$};
\draw[very thick,<-,color=blue] (9,0) -- (9,0.5) node[above, align=center] {delivery $i+1$};
\draw[thick,-To](8,-0.7)--(7,-0.7);
\draw[thick,-To](8,-0.7)--(9,-0.7);
\node [anchor=center]at (8,-1.1){$Y_{(i+1,2)}=Y_{i+1}$};
\draw[thick,->] (-0.2,0) -- (10,0) node[right] {$t$};
\end{tikzpicture}}
\caption{Example of acquisition, transmission and delivery times.}\label{fig:notation}
\end{center}
\end{figure}

Fig.~\ref{fig:notation} shows an example of the notation for acquisition, transmission and delivery times. 
In the delivery $i$ of this example, 
the data unit acquired at $R_{(i,1)}$ is delivered at time $D_i$, with a transmission delay of $Y_{(i,1)}$. 
Then, for delivery $i$, $k_i=1$, $R_{(i,1)}=R_i$ and $Y_{(i,1)}=Y_i$.
In the next delivery, the transmission of the data unit acquired at $R_{(i+1,1)}$ is canceled
$X<Y_{(i+1,1)}$ time units after its transmission, before being delivered. 
Then a second data unit is acquired and transmitted at $R_{(i+1,2)}$, and delivered 
at $D_{i+1}$. For delivery $i+1$, $k_{i+1}=2$, $R_{(i+1,2)}=R_{i+1}$ and  $Y_{(i+1,2)}=Y_{i+1}$.

The AoI is defined as
\begin{equation}
\Delta(t)\triangleq t-\max\{R_i\;:\;D_i\leq t\}.
\end{equation}
For each delivery $i$, the receiver incurs an AoI cost 
over the time interval from the previous delivery time $D_ {i-1}$, to the current delivery time $D_i$.
Let $C_i^{\text{AoI}}$ denote the total AoI cost of delivery $i$,
that is, the total cost in the interval $(D_{i-1},D_i]$ of the AoI perceived by the receiver.
$C_i^{\text{AoI}}$ depends on the function $\Delta(t)$  and the interval $(D_{i-1},D_i]$. Then,
\begin{equation}\label{eq:costAoi}
C_i^{\text {AoI}}= \mathcal{C}(\Delta(t),D_{i-1},D_i),%= f(\Delta(t),D_i,W_i),
\end{equation}
where $\mathcal{C}(\Delta(t),d_1,d_2)$ is a functional which quantifies the cost perceived by the receiver 
due to the information aging in the interval $(d_1,d_2]$ in which the AoI is quantified, 
and $(D_{i-1}$,$D_i]$ is this interval for delivery $i$. 

We generalize the cost definition by considering that each data unit has a transmission cost.
This cost can be used to model any aspect of the transmission that we want to minimize, 
such as the energy consumption in the context of IoT sensor based applications.
Let us call $C_i$ the total cost of delivery $i$, including both the AoI cost and the transmission (or transmissions) cost.
If the cost of transmitting a data unit is a constant value $F$, the total cost of delivery $i$ is
\begin{equation}%\label{eq:transaoicost}
C_i= k_i F+C_i^{\text{AoI}}.
\end{equation}

A common approach to quantify $C_i^{\text{AoI}}$ is through a {\em peak AoI violation cost}~\cite{yates21} , 
assigning it a value of $1$ if the AoI exceeds a specified threshold $A_{\text{th}}$ 
during the interval $(D_{i-1}, D_i]$, and $0$ otherwise.
The threshold  $A_{\text{th}}$ could be any positive value, depending on the application. 
In this case,
\begin{equation}
C_i^{\text{AoI}} =
\begin{cases}
1 & \text{if } \Delta(t)\geq A_{\text{th}},\; \text{for some} \;t\in(D_{i-1},D_i] \\
0 & \text{otherwise},
\end{cases}
\end{equation} 
and, since for $t\in(D_{i-1},D_i]$, $\Delta(t)=t-R_{i-1}$, 
the peak AoI violation cost is
\begin{equation}
C_i^{\text{AoI}} = 
 \mathbbm{1} (D_i-R_{i-1}>A_{\text{th}})
 =\mathbbm{1} (Y_{i-1}+D_i-D_{i-1}>A_{\text{th}}),
\end{equation} 
where $\mathbbm{1}(\cdot)$ is the indicator function.

Another specific definition of the AoI cost, 
introduced in~\cite{sun17}, is as an {\em AoI penalty cost}.
In this case the receiver perceives an instant AoI cost quantified by a non-negative, and non-decreasing
penalty function $p(\Delta(t)),\; p: [0, \infty)\rightarrow [0, \infty)$, satisfying $p(0) = 0$.
For the penalty cost definition,
the AoI cost of delivery $i$ is the accumulated penalty during the interval $(D_{i-1},D_i)$,
\begin{equation}
C_i^{\text{AoI}}= \int_{D_{i-1}}^{D_i}p(\Delta(t)) dt= \int_{Y_{i-1}}^{Y_{i-1}+D_i-D_{i-1}}p(t) dt.
\end{equation}

Some penalty functions were proposed in~\cite{sun17}: the identity function  $p(\Delta(t))=\Delta(t)$, 
the power function $p(\Delta(t))=\Delta(t)^\gamma$, the exponential function $p(\Delta(t))=e^{\gamma\Delta(t)}$, 
and the step function $p(\Delta(t))=\lfloor\gamma\,\Delta(t)\rfloor$. 
Another example of a non linear penalty function is the function  $p(\Delta(t))=\eta \left(e^{\gamma\Delta(t)}-1\right)$, 
adopted in~\cite{niko22,yue22,tang22}.

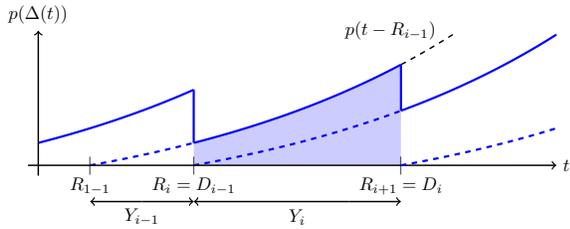
\begin{figure}[tbp]
\begin{center}
\resizebox{0.87\columnwidth}{!}{
\begin{tikzpicture}
\draw[very thick,color=blue]plot[domain=0:3](\x,{exp(0.18*(\x+2))-1});
\fill[fill=blue!20] plot[domain=3:7](\x,{exp(0.18*(\x-1))-1}) |-(3,0);
\draw[very thick,color=blue,dashed]plot[domain=1:3](\x,{exp(0.18*(\x-1))-1});
\draw[very thick,color=blue]plot[domain=3:7](\x,{exp(0.18*(\x-1))-1});
\draw[thick,,dashed]plot[domain=7:8](\x,{exp(0.18*(\x-1))-1});
\node [anchor=center]at (1,0){$|$};
\node [anchor=center]at (1,-0.4){$R_{1-1}$};
\node [anchor=center]at (3,0){$|$};
\node [anchor=center]at (3,-0.4){$R_i=D_{i-1}$};
\draw[thick,-To](2,-0.7)--(1,-0.7);
\draw[thick,-To](2,-0.7)--(3,-0.7);
\node [anchor=center]at (2,-1){$Y_{i-1}$};
\draw[very thick,color=blue,dashed]plot[domain=3:7](\x,{exp(0.18*(\x-3))-1});
\draw[very thick,color=blue]plot[domain=7:10](\x,{exp(0.18*(\x-3))-1});
\node [anchor=center]at (7,0){$|$};
\node [anchor=center]at (7,-0.4){$R_{i+1}=D_i$};
\draw[thick,-To](6,-0.7)--(3,-0.7);
\draw[thick,-To](6,-0.7)--(7,-0.7);
\node [anchor=center]at (5,-1){$Y_i$};
\draw[very thick,color=blue](3,1.46)--(3,0.433);
\draw[very thick,color=blue](7,1.944)--(7,1.054);
\draw[thick,->] (-0.2,0) -- (10,0) node[right] {$t$};
\draw[thick,->] (0,-0.2) -- (0,2.6) node[above] {$p(\Delta(t))$};
\draw[very thick,color=blue,dashed]plot[domain=7:10](\x,{exp(0.18*(\x-7))-1});
\node [anchor=center]at (6.8,2.6){$p(t-R_{i-1})$};
\end{tikzpicture}}
\caption{Example of AoI penalty cost with $R_i=D_{i-1}$.}\label{fig:penalty1}
\end{center}
\end{figure}

An example of AoI penalty cost
is shown in Fig.~\ref{fig:penalty1},
where the solid line represents the evolution of the AoI penalty over time and the shaded area is the AoI cost of delivery $i$.
In this example, no transmissions are canceled and $R_i=D_{i-1}$, that is, 
data acquisitions are made immediately after the previous delivery.

Our goal is to minimize, for any definition of the total cost~$C_i$, the long-term time-average cost, namely
	\begin{equation}\label{eq:beta}
		\beta= %\mathbb{E}[\beta]=
		\lim_{n\rightarrow \infty}\frac{\sum_{i=1}^{n}C_i}{D_n}.
\end{equation}
We assume that the cost rate at any state, $C_i / (D_i - D_{i-1})$, is upper-bounded.
This assumption, together with the previously stated assumption on the sequence of delays (on page~\pageref{MC_assumption}), ensures the existence of the limit above.

We consider two different strategies to reduce $\beta$:
the first, described in Section~\ref{sec:wait} and adopted in~\cite{sun17,tsai23,kam19}, 
is referred to as the  {\em wait strategy}; 
the second, proposed in Section~\ref{sec:discard}, is called the {\em discard strategy}.
These two strategies can provide benefits in different scenarios. For instance, the wait strategy is suitable for scenarios where updates are sent over a reliable channel, whereas the discard strategy is better suited for error-prone or packet-erasure channels.  Although they can be used independently, they are not mutually exclusive and can be applied simultaneously to combine their advantages and achieve a more flexible access strategy.

\subsection{Wait strategy}\label{sec:wait}

\noindent 
In the wait strategy, 
after delivering a data unit, the source waits for an discretionary duration before generating the next one.
When using this strategy, we assume that the communication network has no losses
and the source never cancels a transmission.
Therefore, for all $i$, $k_i=1$, $R_i=R_{(i,1)}$ and $Y_i=Y_{(i,1)}$.
Let $Z_i$ be the wait time for delivery $i$.
After a data unit is delivered at time $D_{i-1}$, the source waits for a duration $Z_i$ before acquiring the next data unit at time $R_i = D_{i-1} + Z_i$, which is then transmitted and, finally, delivered at $D_i$.
Then, $D_i-D_{i-1}=Z_i+Y_i$ and $C_i=F+ C_i^{\text{AoI}}$.
In general, the zero-wait policy $\{Z_i=0,\; \forall i \}$ is not optimal. 
The performance of the zero-wait policy is especially far from the optimum 
when the AoI cost grows quickly with the transmission delay,
the transmission delays are positively correlated over time,
or the transmission delays are highly random~\cite{sun17}.
The sequence of wait times $\{Z_i\}_{i=0}^\infty$ is defined by a causal policy $\pi^\text{w}$, 
from which the wait time $Z_i$ is derived based on the past transmission delays $\{Y_j\}_{j< i}$.
The time-average cost for policy $\pi^\text{w}$ is 
\begin{equation}\label{eq:beta_wait}
 \beta(\pi^\text{w})=
 \lim_{n\rightarrow \infty}\frac{\sum_ {i=1}^n (F+ C_i^{\text{AoI}})}{\sum_ {i=1}^n(Z_i+Y_i)}.
\end{equation}

Let $\Pi^\text{w}$ denote the set of all causal policies satisfying $Z_i \in [0, Z_{\text{max}}],\; \forall i$, 
where an upper bound for the waiting time has been set.  
The optimal value of the time-average cost, if it exists, is 
\begin{equation}\label{eq:optimalbetawait}
 \beta(\pi^\text{w}_{\text{opt}})=\min_{\pi^\text{w}  \in \Pi^\text{w}}\{\beta(\pi^\text{w})\}.
\end{equation}

\begin{figure}[tbp]
\begin{center}
\resizebox{0.87\columnwidth}{!}{
\begin{tikzpicture}
\draw[very thick,color=blue]plot[domain=0:3](\x,{exp(0.18*(\x+2))-1});
\fill[fill=blue!20] plot[domain=3:9](\x,{exp(0.18*(\x-1))-1}) |-(3,0);
\draw[very thick,color=blue,dashed]plot[domain=1:3](\x,{exp(0.18*(\x-1))-1});
\draw[very thick,color=blue]plot[domain=3:9](\x,{exp(0.18*(\x-1))-1});
\node [anchor=center]at (1,0){$|$};
\node [anchor=center]at (1,-0.4){$R_{i-1}$};
\node [anchor=center]at (3,0){$|$};
\node [anchor=center]at (3,-0.4){$D_{i-1}$};
\draw[thick,-To](2,-0.7)--(1,-0.7);
\draw[thick,-To](2,-0.7)--(3,-0.7);
\node [anchor=center]at (2,-1){$Y_{i-1}$};
\draw[very thick,color=blue,dashed]plot[domain=5:9](\x,{exp(0.18*(\x-5))-1});
\draw[very thick,color=blue]plot[domain=9:10](\x,{exp(0.18*(\x-5))-1});
\node [anchor=center]at (5,0){$|$};
\node [anchor=center]at (5,-0.4){$R_i$};
\node [anchor=center]at (9,0){$|$};
\node [anchor=center]at (9,-0.4){$D_i$};
\draw[thick,-To](7,-0.7)--(5,-0.7);
\draw[thick,-To](7,-0.7)--(9,-0.7);
\node [anchor=center]at (7,-1){$Y_i$};
\draw[color=green,thick,-To](4,-0.7)--(3,-0.7);
\draw[color=green,thick,-To](4,-0.7)--(5,-0.7);
\node[anchor=center]at (4,-1){$Z_i$};
\draw[very thick,color=blue](3,1.459)--(3,0.433);
\draw[very thick,color=blue](9,3.221)--(9,1.054);
\draw[thick,->] (-0.2,0) -- (10,0) node[right] {$t$};
\draw[thick,->] (0,-0.2) -- (0,3) node[above] {$p(\Delta(t))$};
\draw[very thick,color=blue,dashed]plot[domain=9:10](\x,{exp(0.18*(\x-9))-1});
\end{tikzpicture}}
\caption{Example of AoI penalty cost with wait strategy.}\label{fig:penalty2}
\end{center}
\end{figure}
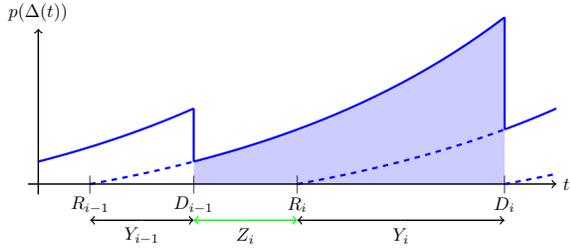

In the case where the cost function is the AoI penalty cost 
and there is no transmission cost ($F=0$), 
the total cost of delivery $i$ under the wait strategy is given by
\begin{equation}\label{eq:costPenaltyF0}
C_i= \int_{Y_{i-1}}^{Y_{i-1}+Z_i+Y_i}p(t) dt,
\end{equation}
which is shown as the shaded area in the example in Figure~\ref{fig:penalty2}.
It has been proven~\cite{sun17} that for this cost function,
if the transmission delays $\{Y_i$\} form a stationary and ergodic Markov process,
there is always a stationary and deterministic optimal policy $\pi_{\text{opt}}^\text{w}$ 
which solves~\eqref{eq:optimalbetawait} and has the form
\begin{equation}\label{eq:waitpolicy}
\pi_{\text{opt}}^\text{w} \triangleq \{Z_i=Z(Y_{i-1})\}_{i=1}^\infty,
\end{equation}
where $Z:\mathbb{R}_{>0}\rightarrow [0,Z_{\text{max}}]$ is a function, 
and $Z_1=0$ is set arbitrarily.
With these assumptions, $\pi_{\text{opt}}^\text{w}$ can be calculated numerically 
with the algorithm described in~\cite{sun17} 
and can be approximated %by the fixed-point $k$-nearest neighbors 
online algorithm described in~\cite{tsai23}.
In all other cases---either when using a strategy other than the wait strategy or a cost function different from the AoI penalty cost given by~\eqref{eq:costPenaltyF0}---the optimal policy, if it exists, cannot be computed or approximated numerically using these algorithms.

In the case where the cost function is the peak AoI violation cost, 
the total cost of delivery $i$ under the wait strategy is 
\begin{equation}
C_i=F+\mathbbm{1}(Y_{i-1}+Z_i+Y_i>A_{\text{th}}),
\end{equation}
and in this case, if $Z_i$ was not limited or if the maximum waiting time $Z_{\text{max}}$ was too large, 
a trivial solution to minimize the long-term time-average of $C_i$
would be to always wait the maximum time.
That is why, with the peak AoI violation cost, the wait strategy only makes sense 
if it is linked to a moderate value of $Z_{\text{max}}$.
This can be interpreted as setting a lower limit to the sampling frequency of the process monitored by the source.

\subsection{Discard strategy}\label{sec:discard}

\noindent 
Previous studies have shown that preemption can improve AoI performance when service-time variability is high (greater than or equal to that of an exponential distribution)~\cite{Kaul12b, Bedewy17, Akar20, Dogan21, Akar21, Fiems23}.
Moreover, the results in~\cite{Prandel24, Qin24, Arafa19, Banerjee24} regarding the structure of the optimal preemption policy support the form of the \textit{discard strategy} we propose here: discarding the ongoing update when its delay exceeds a given threshold.

In the discard strategy, each time a data unit has been delivered, 
the source sets an arbitrary maximum transmission delay for the next data units, until a new data unit is delivered.
Let $X_i$ be the maximum transmission delay set by the source for delivery $i$.
When a data unit is generated and sent at time $R_{(i,1)}$,
the source starts a timer and, if the data unit is not delivered before time $R_{(i,1)}+X_i$, 
the current transmission is canceled and a new data unit is generated and transmitted.
We assume that when the transmission is canceled, the data unit will not be delivered.
This process is repeated until a data unit, say $k_i$, is delivered with a delay $Y_i=Y_{(i,k_i)}<X_i$.
The number of transmissions for delivery $i$ is $k_i\geq 1$, 
then  $D_i-D_{i-1}=(k_i-1) X_i+Y_i$,
and $C_i=k_iF+ C_i^{\text{AoI}}$.

The sequence of maximum transmission delays $\{X_i\}_{i=1}^\infty$ is defined by a causal policy $\pi^\text{d}$, 
from which $X_i$ is derived based on the past transmission delays $\{Y_j\}_{j< i}$. 
The long-term time-average cost for policy $\pi^\text{d}$ is 
\begin{equation}\label{eq:beta_discard}
\beta(\pi^\text{d}) = \lim_{n\rightarrow \infty}\frac{\sum_ {i=1}^n(k_iF+ C_i^{\text{AoI}})}{\sum_ {i=1}^n\left((k_i-1)X_i+Y_i\right)}.
\end{equation}

Let $\Pi^\text{d}$ denote the set of all causal policies satisfying $X_i \in [X_{\text{min}}, X_{\text{max}}],\; \forall i$. 
The optimal value, if it exists, of the time-average cost is 
\begin{equation}\label{eq:optimalbetadisc}
 \beta(\pi_{\text{opt}}^\text{d})= \min_{\pi^\text{d}  \in \Pi^\text{d}} \{\beta(\pi^\text{d}) \},
\end{equation}

\begin{figure}[tbp]
\begin{center}
\resizebox{0.87\columnwidth}{!}{
\begin{tikzpicture}
\draw[very thick,color=blue]plot[domain=0:3](\x,{exp(0.18*(\x+2))-1});
\fill[fill=blue!20] plot[domain=3:8](\x,{exp(0.18*(\x-1))-1}) |-(3,0);
\draw[very thick,color=blue,dashed]plot[domain=1:3](\x,{exp(0.18*(\x-1))-1});
\draw[very thick,color=blue]plot[domain=3:8](\x,{exp(0.18*(\x-1))-1});
\node [anchor=center]at (1,0){$|$};
\node [anchor=center]at (1,-0.4){$R_{i-1}$};
\node [anchor=center]at (3,0){$|$};
\node [anchor=center]at (3,-0.4){$R_{(i,1)}=D_{i-1}$};
\draw[thick,-To](2,-0.7)--(1,-0.7);
\draw[thick,-To](2,-0.7)--(3,-0.7);
\node [anchor=center]at (2,-1){$Y_{i-1}$};
\draw[very thick,color=blue,dashed]plot[domain=3:6.5](\x,{exp(0.18*(\x-3))-1});
\draw[color=red,thick,-To](4,-0.7)--(6.5,-0.7);
\draw[color=red,thick,-To](4,-0.7)--(3,-0.7);
\node [anchor=center]at (4.75,-1){$X_i\; (<Y_{(i,1)})$};
\draw[very thick,color=blue,dashed]plot[domain=6.5:8](\x,{exp(0.18*(\x-6.5))-1});
\draw[very thick,color=blue]plot[domain=8:10](\x,{exp(0.18*(\x-6.5))-1});
\node [anchor=center]at (6.5,0){$|$};
\node [anchor=center]at (6.5,-0.4){$R_{(i,2)}=R_i$};
\node [anchor=center]at (8,0){$|$};
\node [anchor=center]at (8,-0.4){$D_i$};
\draw[thick,-To](7,-0.7)--(6.5,-0.7);
\draw[thick,-To](7,-0.7)--(8,-0.7);
\node [anchor=center]at (7.25,-1){$Y_{(i,2)}=Y_i$};
\draw[very thick,color=blue](3,1.46)--(3,0.433);
\draw[very thick,color=blue,dashed](6.5,0.716)--(6.5,0);
\draw[very thick,color=blue](8,2.525)--(8,0.197);
\draw[thick,->] (-0.2,0) -- (10,0) node[right] {$t$};
\draw[thick,->] (0,-0.2) -- (0,2.5) node[above] {$p(\Delta(t))$};
\draw[very thick,color=blue,dashed]plot[domain=8:10](\x,{exp(0.18*(\x-8))-1});
\end{tikzpicture}}
\caption{Example of AoI penalty cost with discard strategy.}\label{fig:penalty3}
\end{center}
\end{figure}
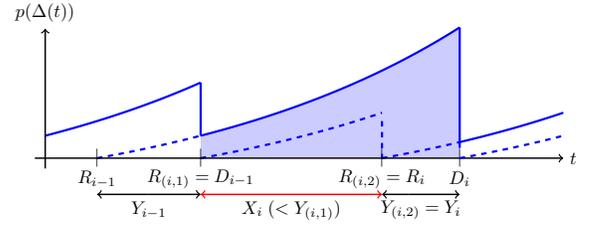

Figure~\ref{fig:penalty3} shows an example, for the AoI penalty cost,
of the evolution of the AoI penalty under the discard strategy.
The shaded area represents the AoI cost, which in this case is 
\begin{equation}% \label{eq:costIntegralDiscard}
C_i^{\text{AoI}}= \int_{Y_{i-1}}^{Y_{i-1}+(k_i-1) X_i+Y_i}p(t) dt.
\end{equation}
 
We have no theoretical evidence, not even for the AoI penalty cost, 
of the existence of an optimal policy which solves~\eqref{eq:optimalbetadisc}, 
nor the characteristics of this policy, but we aim to verify that it is possible, through RL, 
to approximate a stationary and deterministic policy $\pi^\text{d}$, 
\begin{equation}\label{eq:discardpolicy}
\pi^\text{d} \triangleq  \{X_i=X(Y_{i-1})\}_{i=1}^\infty, 
\end{equation} 
with $X:\mathbb{R}_{>0}\rightarrow [X_{\text{min}},X_{\text{max}}]$,
%and $X_1=X_{\text{max}}$, 
and action $X_i$ is given deterministically by the function $X(\cdot)$  applied to the previous state $Y_i$.

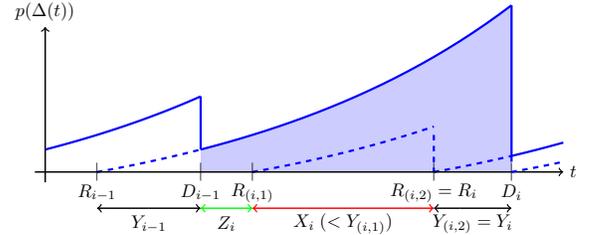
\begin{figure}[tbp]
\begin{center}
\resizebox{0.87\columnwidth}{!}{
\begin{tikzpicture}
\draw[very thick,color=blue]plot[domain=0:3](\x,{exp(0.18*(\x+2))-1});
\fill[fill=blue!20] plot[domain=3:9](\x,{exp(0.18*(\x-1))-1}) |-(3,0);
\draw[very thick,color=blue,dashed]plot[domain=1:3](\x,{exp(0.18*(\x-1))-1});
\draw[very thick,color=blue]plot[domain=3:9](\x,{exp(0.18*(\x-1))-1});
\node [anchor=center]at (1,0){$|$};
\node [anchor=center]at (1,-0.4){$R_{i-1}$};
\node [anchor=center]at (3,0){$|$};
\node [anchor=center]at (3,-0.4){$D_{i-1}$};
\draw[thick,-To](2,-0.7)--(1,-0.7);
\draw[thick,-To](2,-0.7)--(3,-0.7);
\node [anchor=center]at (2,-1){$Y_{i-1}$};
\draw[very thick,color=blue,dashed]plot[domain=4:7.5](\x,{exp(0.18*(\x-4))-1});
\node [anchor=center]at (4,0){$|$};
\node [anchor=center]at (4,-0.4){$R_{(i,1)}$};
\draw[color=red,thick,-To](5,-0.7)--(7.5,-0.7);
\draw[color=red,thick,-To](5,-0.7)--(4,-0.7);
\node [anchor=center]at (5.75,-1){$X_i\; (<Y_{(i,1)})$};
\draw[color=green,thick,-To](3.5,-0.7)--(3,-0.7);
\draw[color=green,thick,-To](3.5,-0.7)--(4,-0.7);
\node[anchor=center]at (3.5,-1){$Z_i$};
\draw[very thick,color=blue,dashed]plot[domain=7.5:9](\x,{exp(0.18*(\x-7.5))-1});
\draw[very thick,color=blue]plot[domain=9:10](\x,{exp(0.18*(\x-7.5))-1});
\node [anchor=center]at (7.5,0){$|$};
\node [anchor=center]at (7.5,-0.4){$R_{(i,2)}=R_i$};
\node [anchor=center]at (9,0){$|$};
\node [anchor=center]at (9,-0.4){$D_i$};
\draw[thick,-To](8,-0.7)--(7.5,-0.7);
\draw[thick,-To](8,-0.7)--(9,-0.7);
\node [anchor=center]at (8.25,-1){$Y_{(i,2)}=Y_i$};
\draw[very thick,color=blue](3,1.46)--(3,0.433);
\draw[very thick,color=blue,dashed](7.5,0.716)--(7.5,0);
\draw[very thick,color=blue](9,3.22)--(9,0.197);
\draw[thick,->] (-0.2,0) -- (10,0) node[right] {$t$};
\draw[thick,->] (0,-0.2) -- (0,2.8) node[above] {$p(\Delta(t))$};
\draw[very thick,color=blue,dashed]plot[domain=9:10](\x,{exp(0.18*(\x-9))-1});
\end{tikzpicture}}
\caption{Example of AoI penalty cost with both wait and discard strategies.}\label{fig:penalty4}
\end{center}
\end{figure}

Both the wait and the discard strategies can be applied simultaneously, 
as shown in an example for the AoI penalty cost in Figure~\ref{fig:penalty4}.

%--------------------------------------------------------------
%--------------------------------------------------------------
\section{Time-average AoI cost reduction through policy gradient RL}\label{sec:algs}
%--------------------------------------------------------------
%--------------------------------------------------------------
\noindent In this section, we propose an RL method to approximate the policies described above. 
The RL algorithms described in this section can be executed online, 
and do not require any knowledge of the distribution of the transmission delays $\{Y_{(i,j)}\}$, 
nor of the function~\eqref{eq:costAoi} that quantifies the AoI cost of each delivery.
It is also valid for any AoI cost function and for total cost functions that include the transmission cost.
Also, the algorithm can be used to implement several strategies simultaneously.

\subsection{Learning setup}

\noindent We propose a policy gradient algorithm,   
which is fully online and incremental.
The algorithm is presented in two versions: one 
that learns an approximation of a state-value function, and a second one that does not need a state-value function.
Each of these versions is applicable to one of the strategies described in Section~\ref{sec:model}.
The algorithm is based on policy gradient methods, 
which learn a parameterized policy to select actions. 
The target optimal policy may be deterministic, 
as in fact it is for the AoI penalty cost with transmission cost $F=0$ and the wait strategy, as proven in~\cite{sun17}.
However, RL requires some amount of exploration, that is, 
for learning to progress it is necessary to take random actions. 
For this reason, a random policy will be learned, 
although in the course of learning the most valuable actions should become increasingly probable.
%Therefore, to approximate a policy $\pi$ it will be necessary to learn the probability density over the set of possible actions conditioned on the system state.
Therefore, to approximate a policy $\pi$, it will be necessary to learn the probability density function over all possible actions, conditioned on the continuous system state.

Assume that the system evolves through a succession of states $S_i$,
and the state changes every time a data unit is successfully delivered at $D_i$.
State changes can occur at any time, 
and the time the system remains in a state is not constant: 
it enters state $S_i$ when a data unit is delivered at time $D_{i-1}$, 
and remains in this state for a time $D_i-D_{i-1}$, 
until the next data unit is successfully delivered.
Let us call step $i$ the period of time during which the system remains in state $S_i$.

The policies that are intended to be learned through RL, 
and in particular the wait policy~\eqref{eq:waitpolicy} and the discard policy~\eqref{eq:discardpolicy}, 
define actions which depend only on the transmission delays.
Accordingly, state $S_i$ is defined by transmission delay of the last delivered data unit,
\begin{equation}
S_i\triangleq Y_{i-1},\; \;Y_{i-1} \in [0, \infty).
\end{equation} 
Thus, the set of states is infinite and continuous.

At any state $S_i$, an action $A_i$ is taken, determined by the policy.
In the wait strategy the action corresponds to the wait time $Z_i=Z(Y_{i-1})$,
while in the discard strategy the action corresponds to the maximum transmission time $X_i=X(Y_{i-1})$.
Like the set of states, the set of actions is a continuous interval: $[0,Z_{\text{max}}]$ or $[X_{\text{min}},X_{\text{max}}]$.

The target policy, $\pi^{w}$ or $\pi^{d}$, will be approximated by a parameterized stochastic policy $\pi(a \cond s, \vec{\theta})$, 
where $a$ is the action, $s$ is the system state and $\vec{\theta} \in \mathbb{R}^d$ is a parameter vector with dimension $d$. 
At step $i$, $s=S_i$, $\vec{\theta}=\vec{\theta}_i$ and 
$\pi(a \cond S_i, \vec{\theta}_i)$ is the probability density function of the random action $A$. 

The policy is learned by updating $\vec{\theta}_i$ at every step. 
Policy gradient methods seek to maximize performance by updating the policy parameters according to:
\begin{equation}\label{eq:policygradient}
\vec{\theta}_{i+1}=\vec{\theta}_i+\alpha_{\theta}\widehat{\nabla J(\vec{\theta}_i)},
\end{equation}
where $\alpha_{\theta}$ is the learning rate, $J(\vec{\theta}_i)$  is the performance measure 
and $\widehat{\nabla J(\vec{\theta}_i)}$ is a stochastic estimate
whose expectation approximates the gradient of $J(\vec{\theta}_i)$ with respect to $\vec{\theta}_i$.
In order to minimize the time-average cost, the performance measure to maximize is
\begin{equation}\label{eq:performeasure}
J(\vec{\theta})=-\beta(\pi(a \cond s, \vec{\theta})),
\end{equation}  
where $\beta(\pi(a \cond s, \vec{\theta}))$ is the long-term time-average cost defined in~\eqref{eq:beta} 
when the policy $\pi(a \cond s, \vec{\theta})$ is followed.
It is clear that this performance measure is a time-average reward,
since $\beta$ is the long-term value of $\beta_n$, 
which defines the time-average cost from $t=0$ up to $t=D_n$.
In RL, the average-reward setting is applicable to undiscounted continuing problems like the one at hand~\cite{mahadevan96}, 
where a  differential return $G_i$ is the difference between
the reward and the average reward, defined as 
\begin{equation}%\label{eq:return}
G_i= \lim_{n\rightarrow\infty}\sum_{j=0}^n (\mathcal{R}_{i+j}-r_{i+j}(\pi)),
\end{equation} 
where $\mathcal{R}_{i}$ is the reward at step $i$, $r_{i}(\pi)$ is the average reward at step $i$ when following policy $\pi$,
and $\mathcal{R}_{i}-r_{i}(\pi)$ is the differential reward at step $i$.

From~\eqref{eq:policygradient}, and applying the policy gradient theorem, the following update can be derived~\cite{sutton18}:
\begin{equation}
\vec{\theta}_{i+1} 
=  \vec{\theta}_i+\alpha_{\theta}\; (G_i-b(Y_{i-1}))\nabla\log\pi(A_i \cond Y_{i-1},\vec{\theta}_i) ,
\end{equation} 
where $b(Y_{i-1})$ is an arbitrary baseline.
By letting  $\delta_i \triangleq G_i-b(Y_{i-1})$, the update for the policy gradient algorithm can be written as
\begin{equation}\label{eq:update}
\vec{\theta}_{i+1} 
=  \vec{\theta}_i+\alpha_{\theta}\; \delta_i\nabla\log\pi(A_i \cond Y_{i-1},\vec{\theta}_i).
\end{equation} 

To maximize the performance measure in~\eqref{eq:performeasure},
the reward at step $i$ has to be $\mathcal{R}_{i}=-C_i$,
and the average reward of step $i$ has to be the negative average cost multiplied by the duration of step $i$:
\begin{equation}
r_{i}(\pi)=-\beta(\pi)(D_i-D_{i-1}).
\end{equation}
Note that $\beta(\pi)(D_i-D_{i-1})$ is the average reward for a step $i$ of duration $D_i-D_{i-1}$,
while $\beta(\pi)$ is the long-term average reward rate.

Actor-critic methods use a learned state-value function $\hat{\nu}_{\pi}(s,\vec{\omega})$,
where $\vec{\omega} \in \mathbb{R}^n$ is a parameter vector of dimension $n$,
to estimate the expected differential return when starting in state $s$ 
and following the policy $\pi$.

The differential return at step $i$ is then estimated as 
\begin{equation}
G_i= -C_i+\beta_i(\pi)(D_i-D_{i-1})+ \hat{\nu}_{\pi}(Y_i,\vec{\omega}_i),
\end{equation}
where $\beta_i(\pi)$ is the estimate of $\beta(\pi)$ at step $i$, 
and $\hat{\nu}_{\pi}(Y_i,\vec{\omega}_i)$ 
is the estimate of the state value of the next state.

The parameter vector $\vec{\omega}$ is updated at each step 
via stochastic gradient descent (SGD):
\begin{equation}
\vec{\omega}_{i+1}= \vec{\omega}_i + \alpha_{\omega}\; \delta_i\; \nabla\hat\nu(Y_{i-1},\vec{\omega}_i),
\end{equation} 
where $\alpha_{\omega}$ is the learning rate.
	
The baseline $b(Y_{i-1})$  can be any function, even a random variable, 
that does not vary with the action $a$.
Depending on the strategy used, the estimate
$\hat{\nu}_{\pi}(s_i,\vec{\omega}_i)$ may or may not vary with $a$.
For example, with the wait strategy, $\hat{\nu}_{\pi}(s_i,\vec{\omega}_i)$ does not vary with the action, 
because $S_i$ does not depend on $A_i$, that is, the delay $Y_i$ does not depend on action $Z_i$.
Therefore, for the wait strategy  we can set $b(Y_{i-1})=-\hat{\nu}_{\pi}(Y_i,\vec{\omega}_i)$, and $\delta_i$ boils down to 
\begin{equation}\label{eq:deltapg}
\delta_i=  -C_i+\beta_i(\pi)(D_i-D_{i-1}).
\end{equation} 

For those cases in which~\eqref{eq:deltapg} can be used,
as is the case of the wait strategy, 
we propose an algorithm in Section~\ref{sec:alg12} 
that does not need to learn an approximation of any state-value function. 
This algorithm belongs to the general category of policy gradient algorithms.

For those cases in which  $\hat{\nu}_{\pi}(s_i,\vec{\omega}_i)$ varies with $a$,
as is the case of the discard strategy, 
we propose and actor-critic algorithm, also in Section~\ref{sec:alg12},
for which the baseline  $b(Y_{i-1})$  is the estimate of the state-value of the current state, 
$\hat{\nu}_{\pi}(Y_{i-1},\vec{\omega}_i)$, and $\delta_i$ is
\begin{equation}
\delta_i
=  -C_i+\beta_i(\pi)(D_i-D_{i-1})+ \hat{\nu}_{\pi}(Y_i,\vec{\omega}_i)-\hat{\nu}_{\pi}(Y_{i-1},\vec{\omega}_i).
\end{equation} 

\subsection{Algorithms}\label{sec:alg12}

\noindent Algorithm~\ref{alg:1} implements the learning setup for the wait strategy. 
The algorithm is run in a continuous loop. 
The current step starts at time $D$, which is the delivery time of the last delivered data unit.
At the current step, the system state is $Y$, which is the transmission delay of the last delivered data unit.
At the beginning of the step, the source generates a random wait time $Z$ 
with the distribution for state $Y$ given by the parameterized policy $\pi$  (line 7).
Then it waits a time $Z$, gets the next data unit, transmits it and waits for the feedback.
When the next data unit is delivered (line 11), 
the source observes its transmission delay $Y_\text{next}$  and the total cost of the current delivery $C_u$.
From this information, $\delta$ is calculated  (line 14) according to~\eqref{eq:deltapg}
and the policy parameters are updated (line 15), according to~\eqref{eq:update}.
Then, the delivery time $D$ is updated,
$Y_\text{next}$ becomes the current state, and the next step starts.
At each step, the estimate of the average cost used to calculate $\delta$
is the quotient between the total cost from the start to the current step $C$ 
and the initial time of the current step $D$.

\newcommand\mycommfont[1]{\footnotesize\ttfamily\textcolor{blue}{#1}}
\SetCommentSty{mycommfont}
\SetKwInput{KwInput}{Input}                % Set the Input
\SetKwInput{KwOutput}{Output}              % set the Output

\begin{algorithm}[tbp]
\DontPrintSemicolon
set $\alpha_{\theta}$  \tcp*{learning rate}    
set $F$        \tcp*{transmission cost}    
$\vec{\theta}\leftarrow\vec{0}$ \tcp*{policy  parameters}     
$D\leftarrow 1$ \tcp*{delivery time} 
$C\leftarrow 0$  \tcp*{cost from the start}
\Repeat{forever}{  
	$Z\backsim \pi(\cdot \cond Y,\vec{\theta})$     \tcp*{generate random $Z$}      
	{\bf wait} $Z$ \\
	get data and transmit \\
	{\bf wait} {\em feedback} \\
	observe $Y_{\text{next}}$ and $C_u$ \tcp*{delay, delivery cost}
	$W\leftarrow Z+Y_{\text{next}}$ \tcp*{step duration}      
    $C\leftarrow C + F+C_u$ \\
    $ \delta=-F-C_u+ WC/D$  \\
	$\vec{\theta} \leftarrow \vec{\theta} + \alpha_{\theta}\; \delta\; 
	\nabla\log\pi(Z \cond Y,	\vec{\theta})$ \tcp*{update $\vec{\theta}$} 
	$D\leftarrow D+W$ \tcp*{next delivery time} 
	$Y\leftarrow Y_{\text{next}}$ \tcp*{next state}
}
\caption{Policy gradient algorithm for the wait strategy.}\label{alg:1}
\end{algorithm}

Algorithm~\ref{alg:2} implements the learning setup described above for the discard strategy. 
At each step, the source generates a random maximum transmission delay $X$ 
with the distribution given by a parameterized policy $\pi$ (line 8).
Then it gets a data unit, transmits it and waits a maximum time $X$
or until the data unit is delivered and the feedback is received. 
Every time that a data unit is not delivered before time $X$ (line 13), 
the transmission is canceled,
a new data unit is generated and transmitted (line 11)
and the wait for a maximum time $X$ or the data delivery starts again.
When a data unit is delivered, the reward average rate is estimated, $\delta$ is calculated 
and the policy parameters are updated.

\begin{algorithm}[tbp]
\DontPrintSemicolon
set $\alpha_{\theta}, \alpha_{\omega} $  \tcp*{learning rates}    
set $F$        \tcp*{transmission cost}
$\vec{\theta}\leftarrow\vec{0}$ \tcp*{policy  parameters}     
$\vec{\omega}\leftarrow\vec{0}$ \tcp*{state-value  parameters}   
$D\leftarrow 1$ \tcp*{delivery time} 
$C\leftarrow 0$  \tcp*{cost from the start}
\Repeat{forever}{  
	$X\backsim \pi(\cdot \cond Y,\vec{\theta})$     \tcp*{generate random $X$}     
	$k\leftarrow 1$ \tcp*{number of transmissions} 
	\Repeat{feedback}{
		 get data and transmit \\
		{\bf wait} {\bf (}$X$ {\bf or} {\em feedback} {\bf )} \\
	\If{X}{      
		cancel transmission  \tcp*{$X$ expired}  
		$k\leftarrow k+1$ \\
		} 
	}	
	observe $Y_{\text{next}}$ and $C_u$ \tcp*{delay, delivery cost}
	$W\leftarrow (k-1)X+Y_{\text{next}}$  \tcp*{step duration}
    $C\leftarrow C + kF+C_u$ \\
    $ \delta=-kF-C_u+ WC/D+ \hat\nu(Y_{\text{next}},\vec{\omega})-\hat\nu(Y,	\vec{\omega})$  \\
	$\vec{\theta} \leftarrow \vec{\theta} + \alpha_{\theta}\; \delta\; 
	\nabla\log\pi(X \cond Y, 	\vec{\theta})$ \tcp*{update $\vec{\theta}$}
	$\vec{\omega} \leftarrow \vec{\omega} + \alpha_{\omega}\; \delta\; \nabla\hat\nu(Y,
	\vec{\omega})$ \tcp*{update $\vec{\omega}$}
	$D\leftarrow D+W$ \tcp*{next delivery time} 
	$Y\leftarrow Y_{\text{next}}$ \tcp*{next state}
}
\caption{Actor-critic algorithm for the discard strategy.}\label{alg:2}
\end{algorithm}

Algorithms~\ref{alg:1} and~\ref{alg:2} can be modified, or even combined, 
to implement simultaneously more than one strategy, without increasing noticeably the computational cost.
As an example, Algorithm~\ref{alg:3} simultaneously implements the two previous strategies: wait and discard.
For each of the two strategies, the algorithm learns a separate policy, with parameters $\theta_w$ and $\theta_d$, respectively.
Since one of these strategies affects the state transition probabilities, a state-value function has to be learned also.
At each step, both strategies are implemented and the respective policies are updated.
The state-value is also used and updated at any step. 
Similarly, this approach could be applied to other strategies or their combinations by using a separate policy for each strategy and implementing the strategies at every step of the algorithm.

The learning of each policy will be driven by the use of all the strategies combined, 
which means that the policy learned for each strategy while other strategies 
are  simultaneously implemented will be different from the policy learned when only this strategy is implemented.

\begin{algorithm}[tbp]
\DontPrintSemicolon
set $\alpha_{\theta}, \alpha_{\omega} $  \tcp*{learning rates}    
set $F$        \tcp*{transmission cost}
$\vec{\theta_w}\leftarrow\vec{0}$ \tcp*{wait policy  parameters}  
$\vec{\theta_d}\leftarrow\vec{0}$ \tcp*{discard policy  parameters}   
$\vec{\omega}\leftarrow\vec{0}$ \tcp*{state-value  parameters}   
$D\leftarrow 1$ \tcp*{delivery time} 
$C\leftarrow 0$  \tcp*{cost from the start}
\Repeat{forever}{  
    $Z\backsim \pi(\cdot \cond Y,\vec{\theta_w})$    \tcp*{generate random $Z$}    	
	$X\backsim \pi(\cdot \cond Y,\vec{\theta_d})$    \tcp*{generate random $X$}     
	$k\leftarrow 1$ \tcp*{number of transmissions} 
	{\bf wait} $Z$ \\
	\Repeat{feedback}{
		 get data and transmit \\
		{\bf wait} {\bf (}$X$ {\bf or} {\em feedback} {\bf )} \\
	\If{X}{      
		cancel transmission  \tcp*{$X$ expired}  
		$k\leftarrow k+1$ \\
		} 
	}	
	observe $Y_{\text{next}}$ and $C_u$ \tcp*{delay, delivery cost}
	$W\leftarrow Z+(k-1)X+Y_{\text{next}}$  \tcp*{step duration}
    $C\leftarrow C + kF+C_u$ \\
    $ \delta=-kF-C_u+ WC/D+ \hat\nu(Y_{\text{next}},\vec{\omega})-\hat\nu(Y,	\vec{\omega})$  \\
	$\vec{\theta_w} \leftarrow \vec{\theta_w} + \alpha_{\theta}\; \delta\; 
	\nabla\log\pi(Z \cond Y,\vec{\theta_w})$ \tcp*{\hspace{-4pt}update $\vec{\theta_w}$}
	$\vec{\theta_d} \leftarrow \vec{\theta_d} + \alpha_{\theta}\; \delta\; 
	\nabla\log\pi(X \cond Y,\vec{\theta_d})$ \tcp*{update $\vec{\theta_d}$}
	$\vec{\omega} \leftarrow \vec{\omega} + \alpha_{\omega}\; \delta\; \nabla\hat\nu(Y,
	\vec{\omega})$ \tcp*{update $\vec{\omega}$}	
	$D\leftarrow D+W$ \tcp*{next delivery time} 
	$Y\leftarrow Y_{\text{next}}$ \tcp*{next state}}
\caption{Algorithm for the simultaneous use of wait and discard strategies.}\label{alg:3}
\end{algorithm}

In approaches other than policy gradient methods, 
	each additional strategy introduces a new dimension to the learned action-value function, 
	leading to an exponential increase in the complexity of algorithms such as SARSA. 
	In contrast, Algorithms~\ref{alg:1} and~\ref{alg:2} do not rely on an action-value function. 
	When multiple strategies are learned simultaneously, each policy selects actions independently, 
	without referencing the others, and each is updated separately. 
	As a result, the complexity of Algorithms~\ref{alg:1} and~\ref{alg:2} grows linearly with the number of simultaneously implemented strategies.

\subsection{Function approximations}

\noindent In Algorithm~\ref{alg:2}, the policy $\pi(X \cond Y,\vec{\theta})$ to be learned 
is a continuous probability distribution whose support is the interval $(X_{\text{min}},X_{\text{max}})$.
Furthermore, since it must be able to approximate a deterministic policy,
the distribution must be unimodal, 
and it should be possible to adjust its variance.
A distribution that meets these requirements 
can be obtained from a lognormal distribution $X'$, 
$\log X'\backsim \mathcal{N}(\mu,\sigma^2)$, transformed as
\begin{equation}\label{eq:lognormalx}
X=X_{\text{max}}-\frac{X_{\text{max}}-X_{\text{min}}}{1+X'}.
\end{equation}

Similarly, in Algorithm~\ref{alg:1}, the policy $\pi(Z \cond Y,\vec{\theta})$ to be learned 
is a continuous probability distribution with support  $(0,Z_{\text{max}})$
obtained from a lognormal distribution $Z'$, 
$\log Z'\backsim \mathcal{N}(\mu,\sigma^2)$, transformed as in~\eqref{eq:lognormalx}, with $Z_{\text{min}}=0$.

In both cases, $\sigma$ 
is a fixed parameter that can be tuned to control the dispersion of the distribution, 
whereas $\mu$ is defined as a parameterized function of $Y$ and is adjusted during the learning process, 
so that $\pi(Z \mid Y, \vec{\theta})$ and $\pi(X \mid Y, \vec{\theta})$ approximate the optimal policy.
We have used a linear approximation for $\mu$, so that each algorithm aims to learn a function
\begin{equation}
\mu(Y,\vec{\theta})=\vec{\theta}^T \vec{f}(Y)=\sum_{k=1}^d \theta^k f_k(Y),
\end{equation}
where $\theta^k$ is the $k$-th component of the parameter vector $\vec{\theta}$ 
and $\vec{f}(Y)$ is a feature vector representing state $Y$. 
The component $f_k(Y)$ of $\vec{f}(Y)$ is the value of a function $f_k : [0,Y_{\text{max}}]\rightarrow \mathbb{R}$, 
where $[0,Y_{\text{max}}]$ is the interval of the state set for which the policy is learned.
$Y_{\text{max}}$ is chosen so that values of the transmission delay $Y>Y_{\text{max}}$ are very unlikely. 
Outside this interval, the policy takes a fixed value
(for example, $Z(Y)=0,\; \forall \; Y\geq Y_{\text{max}}$,
and $X(Y)=X_{\text{max}},\; \forall \; Y\geq Y_{\text{max}}$).

For the linear approximation, we have implemented a Fourier series with only cosine functions 
and fundamental pulsation $\pi/Y_{\text{max}}$.
The feature functions are the cosine functions 
and the Fourier's series is
\begin{equation}\label{eq:fourier}
\mu(Y,\vec{\theta})=\sum_{k=0}^{d-1} \theta^k\cos\frac{k\pi Y}{Y_{\text{max}}},
\end{equation}
which is a periodic function with period $2Y_{\text{max}}$. 
The parameter vector $\vec{\theta}$ must be learned to approximate $\mu(Y,\vec{\theta})$ 
in the half-period $[0,Y_{\text{max}}]$.

The probability density function of $X$, from
the lognormal probability density function and~\eqref{eq:lognormalx}, is
\begin{equation}\label{eq:policypdf}
\pi(X \cond Y,\vec{\theta})= \frac{(X_{\text{max}}-X)}{\sigma (X-X_{\text{min}}) \sqrt{2\pi}}e^{-\frac{\left(\log\frac{X-X_{\text{min}}}{X_{\text{max}}-X}-\mu(Y,\vec{\theta})\right)^2}{2\sigma^2}}.
\end{equation} 

By taking the partial derivatives of the logarithm of $\pi(X \cond Y,\vec{\theta})$ 
with respect to $\theta^k,\; k=0, \dots, d-1$, 
we obtain the eligibility vector used in~\eqref{eq:update}  to update the policy:
\begin{equation}\label{eq:eligibvector}
\nabla\log\pi(X \cond Y,\vec{\theta})=
\frac{\log \frac{X-X_{\text{min}}}{X_{\text{max}}-X}-\mu(Y,\vec{\theta})}{\sigma^2}
\nabla \mu(Y,\vec{\theta}),
\end{equation} 
where
\begin{equation}\label{eq:fouriergradient}
\nabla \mu(Y,\vec{\theta})= 
\left(1,\cos\frac{\pi Y}{Y_{\text{max}}},\dots, \cos\frac{(d-1)\pi Y}{Y_{\text{max}}} \right).
\end{equation} 

From this, the policy update implemented in Algorithm~\ref{alg:2} (line 22) is
\begin{equation}
\theta_{i+1}^k =  \theta_i^k +\alpha_{\theta}\; \delta_i
\frac{\log \frac{X_i-X_{\text{min}}}{X_{\text{max}}-X_i}-\mu(Y,\vec{\theta})}{\sigma^2}
\cos \frac{k\pi Y}{Y_{\text{max}}},
\end{equation} 
for $k=0\dots d-1$,
where $\theta_i^k$ is the $k$-th component of the parameter vector at step $i$, $\vec{\theta}_i$.
The update in Algorithm~\ref{alg:1} (line 15) is obtained in an analogous way.
%Similarly, the update in Algorithm~\ref{alg:1} (line 15) is:
%\begin{multline}
%\theta_{i+1}^k = \theta_i^k + \alpha_{\theta}\; \delta_i
%\frac{\log \frac{Z_i}{Z_{\text{max}}-Z_i}-\mu(Y,\vec{\theta})}{\sigma^2}
%\cos\frac{k\pi Y}{Y_{\text{max}}}. \\ \nonumber
%k=0\dots d-1.
%\end{multline} 

The state-value function $\hat\nu(Y,\vec{\omega})$ has also been linearly approximated 
by a Fourier's series as in~\eqref{eq:fourier} with $n$ components
\begin{equation}
\hat\nu(Y,\vec{\omega})=\vec{\omega}^T \vec{f}(Y)=\sum_{k=1}^n \omega^k f_k(Y),
\end{equation}
where $\omega^k$ is the $k$-th component of the parameter vector $\vec{\omega}$ and $\vec{f}(Y)$ 
is the feature vector,
with component $f_k(Y),\; f_k : [0,Y_{\text{max}}]\rightarrow \mathbb{R}$,
\begin{equation}
 f_k(Y)=\cos\frac{k\pi Y}{Y_{\text{max}}},
\end{equation}
and the gradient $\nabla\hat\nu(Y,\vec{\omega})$ is a vector of harmonic functions,
as in~\eqref{eq:fouriergradient}, with $n$ components.

%--------------------------------------------------------------
%--------------------------------------------------------------
\section{Results and discussion}\label{sec:results}
%--------------------------------------------------------------
%--------------------------------------------------------------
\noindent In this section, we evaluate the performance of Algorithm~\ref{alg:1}, the wait strategy, and Algorithm~\ref{alg:2}, the discard strategy, through an extensive series of simulations, considering a comprehensive set of penalty and cost functions, 
and a broad set of benchmarks. 
Specifically, we compare the
obtained policies ($Z(Y_i)$ and $X(Y_i)$) and the resulting time-average performance ($\beta$)
of our algorithms with the following benchmarks.
\begin{enumerate}
    \item \emph{State-of-the-art methods:} Two methods recently proposed in the literature, denoted as \emph{Benchmark~1}~\cite{kam19} and \emph{Benchmark~2}~\cite{tsai23} are considered. 
%    Benchmark~1 is based on SARSA~\cite{kam19} and 
%    Benchmark~2 is based on K-nearest neighbors (KNN)~\cite{tsai23}.
     Like our algorithms, these methods operate online and require no prior knowledge of the channel model or the cost function.
    \item \emph{Null policies:} The null policy for the wait strategy is the \textit{zero-wait} policy with $Z_i = 0$ for all $i$ and the null policy for the discard strategy is the \textit{maximum-delay} policy with $X_i=Y_\text{max}$ for all $i$.
    \item \emph{Optimal policy:} The optimal policy can be obtained in limited cases using the offline method from~\cite{sun17}, which assumes perfect knowledge of the channel model and the cost function. Moreover, the optimal policy can be calculated numerically in scenarios with greatly simplified channel models. 
\end{enumerate}
To cover the wide range of scenarios considered, this section is organized as follows.  Section~\ref{sec:wait_specific_cost} presents the results for Algorithm~\ref{alg:1} when the cost function adheres to~\eqref{eq:costPenaltyF0}. 
Section~\ref{sec:wait_general_cost} extends this analysis to general cost functions that do not conform to~\eqref{eq:costPenaltyF0}. 
Section~\ref{sec:discard_results} examines the results for Algorithm~\ref{alg:2},
while Section~\ref{sec:twostrategies} explores the performance of the combined application of wait and discard strategies.

Although our algorithms and Benchmark~1 are applicable across all scenarios,
Benchmark~2 and the optimal solution are limited to the wait strategy in cases where the cost function adheres to~\eqref{eq:costPenaltyF0}.
Consequently, for the scenarios considered in Sections~\ref{sec:wait_general_cost} through~\ref{sec:twostrategies}, 
it remains unclear how close our algorithms are to the optimum: 
to the best of our knowledge, no method exists for computing the optimal policy in these cases. 
To address this limitation, 
Section~\ref{sec:simplified_channel_model} provides results using a simplified channel model, enabling the computation of the optimal policy via conventional discrete-MDP methods. 
Lastly, Section~\ref{sec:comp_cost} evaluates the computational cost of all methods used in this paper.

Except for Section~\ref{sec:simplified_channel_model}, we model the transmission delay $Y_i$ as a log-normally distributed Markov process, 
given by $Y_i=e^{\sigma_d S_i}/\mathbb{E}[e^{\sigma_d S_i}]$, 
where $S_i$ represents the state of a Markov process with an infinite and continuous state space and where we set $\sigma_d=1.5$. The state transitions are generated using the autoregressive formula $S_{i+1}=\eta S_i+\sqrt{1-\eta^2}N_i$, where $\{N_i\}_{i=1\dotsc,\infty}$ is a set of i.i.d. random variables with a standard normal distribution $\mathcal{N}(0,1)$. The process $Y_i$ is stationary, with a correlation coefficient between $Y_i$ and $Y_{i+1}$ given by $\rho=(e^\eta-1)/(e-1)$.

The results are obtained by simulation, 
by calculating the average over 100 simulation runs with $10^6$ time units per run. 
For Benchmark~1, the learning rate has been set to $\alpha=10^{-2}$.
For both Algorithm~\ref{alg:1} and Algorithm~\ref{alg:2},
the learning rate is $\alpha_{\theta}=10^{-4}$,
the standard deviation of the stochastic policies~\eqref{eq:policypdf}
%(see ~\eqref{Theeq:lognormalx} and the preceding text)
is $\sigma=0.5$,
and the dimensions of the parameter vectors that approximate the policy and the state-value function are $d=10$ and $n=10$ respectively. 

\subsection{Wait strategy with AoI penalty and no transmission cost}\label{sec:wait_specific_cost}

\begin{figure}[tbp]
\centering
%\subfloat[]{\includegraphics[width=\columnwidth]{aoiIdRho00}}\\
\subfloat[]{\includegraphics[width=\columnwidth]{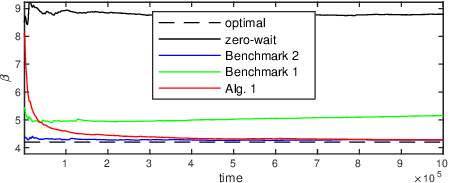}} \\
\subfloat[]{\includegraphics[width=\columnwidth]{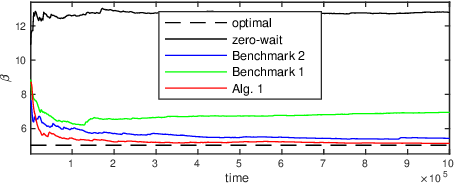}}
\caption{Time-average costs for wait strategy and AoI penalty cost with identity function. (a) $\rho=0.5$. (b) $\rho=0.9$. } 
\label{fig:idrho}
\end{figure}

\begin{figure}[tbp]
\begin{center}
\includegraphics[width=\columnwidth]{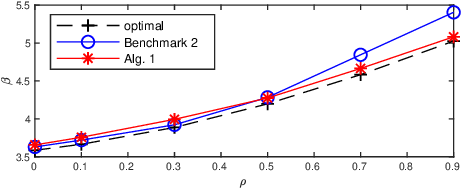}
\caption{Performance comparison of Algorithm~\ref{alg:1} and Benchmark~2.}\label{fig:aoirho}
\end{center}
\end{figure}

\begin{figure}[tbp]
\begin{center}
\includegraphics[width=\columnwidth]{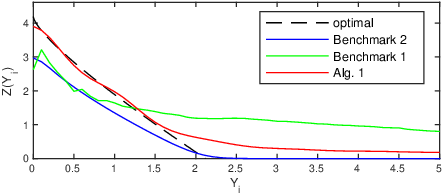}\\[-0.5em]
\caption{Wait policies for AoI penalty cost with identity function, $\rho=0.5$.}\vspace{-10pt}
\label{fig:policyrho05}
\end{center}
\end{figure}

\begin{figure}[tbp]
\begin{center}
\includegraphics[width=\columnwidth]{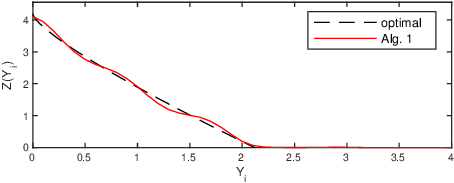}\\[-0.5em]
\caption{Wait policy learned after $10^8$ steps with Algorithm~\ref{alg:1}.}
\label{fig:polconvidrho05}
\end{center}
\end{figure}

\begin{figure}[tbp]
\centering
\subfloat[]{\includegraphics[width=\columnwidth]{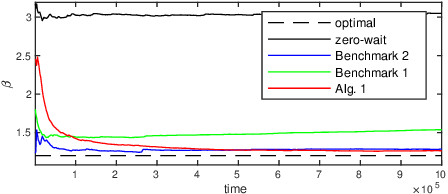}}\\
\subfloat[]{\includegraphics[width=\columnwidth]{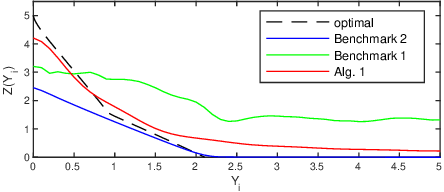}}
\caption{(a) Time-average costs and (b) policies for wait strategy and AoI penalty cost  with step function with $\gamma=0.4$.}
\label{fig:step04} 
\end{figure}

\noindent This section presents the time-average costs and resulting policies with the wait strategy (i.e., Algorithm~\ref{alg:1}), both Benchmark~1 and~2,  the zero-wait policy, and the optimal policy with the cost function defined in~\eqref{eq:costPenaltyF0} for two penalty functions of the AoI: 
identity function $p(\Delta(t))=\Delta(t)$, shown in Figures~\ref{fig:idrho}--\ref{fig:polconvidrho05}, and step function $p(\Delta(t))=\lfloor \gamma\,\Delta(t)\rfloor$, shown in Figure~\ref{fig:step04}.

Figure~\ref{fig:idrho} illustrates the evolution of the time-average cost 
for two levels of  correlation  between $Y_{i-1}$ and $Y_i$:   $\rho=0.5$ and $\rho=0.9$.
In both cases, the zero-wait policy performs significantly worse, 
with only Benchmark~2 and Algorithm~\ref{alg:1} approaching the optimum. 
Benchmark~1, in contrast, exhibits convergence issues and performs significantly worse than the optimal policy.

Figure~\ref{fig:aoirho} shows a comparison of the time-average costs obtained
by the two online algorithms with the best performance, 
which are  Algorithm~\ref{alg:1} and Benchmark~2, for a wider range of $\rho$ values. 
While both algorithms obtain a result quite close to the optimum, Benchmark~2 performs slightly better 
with lower values of $\rho$, when consecutive values of $Y_i$ are almost independent. However,  it is outperformed by Algorithm~\ref{alg:1} when consecutive values of $Y_i$ are highly correlated.
This indicates that Algorithm~\ref{alg:1} is more capable of capturing the memory of the transmission delay process.

Figure~\ref{fig:policyrho05} shows the wait policies learned and used 
by the different algorithms in the previous experiment,
compared with the optimal policy,
for $\rho=0.5$. % value in the center of the range shown in Figure~\ref{fig:aoirho}.
The resulting policies are measured by registering the values of $Z_{i+1}=Z(Y_i)$ applied 
for each interval of values of $Y_i$.
The policies learned by Benchmark~1 are far from the optimum, which explains its poor performance.
In contrast, 
the policy learned by Algorithm~\ref{alg:1} aligns more closely with the optimal policy 
than Benchmark~2 for values of $Y_{i-1}$ where the optimal policy is $Z_i > 0$. 
However, it incurs an error for $Y_{i-1} > 2$, which does not occur for Benchmark~2, where the optimal policy is $Z_i = 0$. 
This is due to the fact that Algorithm~\ref{alg:1} learns a stochastic policy 
and it generates random values $Z_i>0$ even when $Z_i$ should be $0$,
while all values of $Z_i$ in the optimal policy are deterministic.
Nevertheless, this error decreases as Algorithm~\ref{alg:1} runs longer and 
these values approach $0$.
To find out how accurate the policy learned by  Algorithm~\ref{alg:1} is in the long-term,
we let the algorithm run for a longer time.
Figure~\ref{fig:polconvidrho05} plots the policy learned after $10^8$ steps 
alongside the optimal policy, 
demonstrating that the learned policy converges to it with high accuracy.

Figure~\ref{fig:step04} shows the evolution of the time-average cost 
and the policies with the penalty function $p(t)=\lfloor \gamma\,t\rfloor\;$ with $\gamma=0.4$, for the channel model with $\rho=0.5$.
Also in this case, both Benchmark~2 and  Algorithm~\ref{alg:1} approach the optimal time-average cost,
while Benchmark~1 deviates significantly from it.
For the step function, the optimal wait policy 
is not a straight line, but is composed of two sections and,
of the three policies shown, the only one that closely matches to the shape of the optimal policy 
is the one learned by  Algorithm~\ref{alg:1}.

\subsection{Wait strategy and general cost function}\label{sec:wait_general_cost}

\noindent
Unlike Sec.~\ref{sec:wait_specific_cost}, 
we now consider two examples of cost functions that do not follow the specific form of~\eqref{eq:costPenaltyF0}. 
Consequently, the optimal policy cannot be calculated, and Benchmark~2 is not applicable either.
Therefore, the following figures show results only for Benchmark~1 and Algorithm~\ref{alg:1}.
The two cost functions used are:
\begin{itemize}
	\item AoI penalty cost with identity penalty function and transmission cost $F>0$,
		\begin{equation}\label{eq:idcostwait}
			C_i=F+\int_{Y_{i-1}}^{Y_{i-1}+Z_i+Y_i}t\; dt,
		\end{equation}
	\item Peak AoI violation,
		\begin{equation}\label{eq:thrcostwaitF0}
			C_i=\mathbbm{1}(Y_{i-1}+Z_i+Y_i>A_{\text{th}}).
		\end{equation}
\end{itemize}

\begin{figure}[tbp]
\centering
\subfloat[]{\includegraphics[width=\columnwidth]{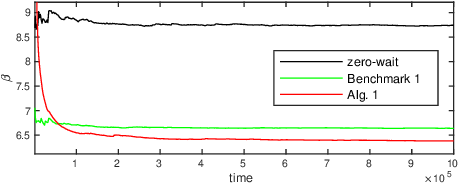}}\\
\subfloat[]{\includegraphics[width=\columnwidth]{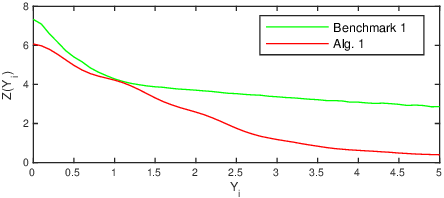}}
\caption{(a) Time-average costs and (b) policies for the wait strategy and the sum of the AoI penalty cost (defined by the identity function) and a transmission cost with $F=10$.}
\label{fig:cost10}
\end{figure}

Figure~\ref{fig:cost10} shows the evolution of the time-average costs and the policies 
for the cost function given by~\eqref{eq:idcostwait} with $F=10$ and the channel model with $\rho=0.5$.
It can be seen that both algorithms are capable of significantly reducing the time-average cost, 
with a better result from  Algorithm~\ref{alg:1}.
Although it is not possible to compare the learned policies with the optimal policy, 
it is observed that the policy learned by Algorithm~\ref{alg:1} is qualitatively similar 
to the optimal policy in the same case with $F=0$ (Figure~\ref{fig:polconvidrho05}).

\begin{figure}[tbp]
\centering
\subfloat[]{\includegraphics[width=\columnwidth]{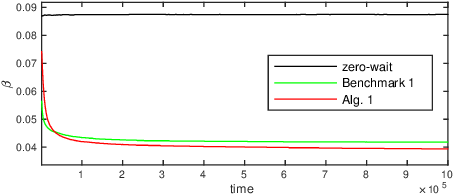}}\\
\subfloat[]{\includegraphics[width=\columnwidth]{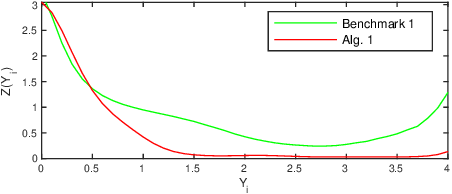}}
\caption{(a) Time-average costs and (b) policies for wait strategy and peak AoI violation cost with $A_{\text{th}}=5$.}\label{fig:thin5}
\end{figure}

Figure~\ref{fig:thin5} shows the evolution of the time-average costs and the policies 
for the cost function given by~\eqref{eq:thrcostwaitF0} 
with $A_{\text{th}}=5$ and the channel model with $\rho=0.5$.
In this case, Benchmark~1 performs better than in previous scenarios,
but it is still outperformed by Algorithm~\ref{alg:1}.

The wait policies learned by both algorithms follow a pattern similar
to that of the optimal policies of the previous cases.

\subsection{Discard strategy}\label{sec:discard_results}

\noindent This section presents results for the discard strategy.
The time-average costs obtained with Algorithm~\ref{alg:2} are compared with the results of Benchmark~1
and of the null (i.e., maximum-delay) policy where
$X_i=Y_{\text{max}}$ for all $i$.
These results are obtained for a cost function which includes an AoI penalty cost
with the identity penalty function and a transmission cost. For the discard strategy, this function is, 
\begin{equation}\label{eq:idcostdiscF}
C_i=k_iF+\int_{Y_{i-1}}^{Y_{i-1}+(k_i-1)X_i+Y_i}t\; dt.
\end{equation}

\begin{figure}[tbp]
\begin{center}
\includegraphics[width=\columnwidth]{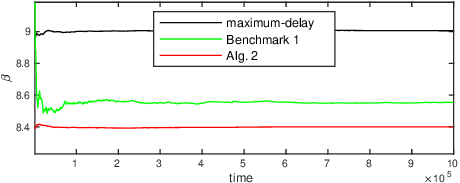}\\[-0.5em]
\caption{Time-average costs for the discard strategy and the sum of the AoI penalty cost (defined by the identity function) and a transmission cost with $F=4$. }\label{fig:costdiscard}\vspace{-10pt}
\end{center}
\end{figure}

Figure~\ref{fig:costdiscard} shows the evolution of the time-average costs
for $F=4$ and the log-normal channel model with $\rho=0.5$.
The learned discard policies are bounded by $X_{\text{min}}\leq X \leq X_{\text{max}}$, 
with $X_{\text{min}}=2$ and $X_{\text{max}}=Y_{\text{max}}=10$.
From this result, we see that, with this policy, 
the behavior in terms of performance is quite similar to that with the wait strategy.
Both algorithms are capable of significantly reducing the time-average cost, 
but Algorithm~\ref{alg:2} achieves the best result,
although we cannot assess how close it is to the optimum.

\begin{figure}[tbp]
\begin{center}
\includegraphics[width=\columnwidth]{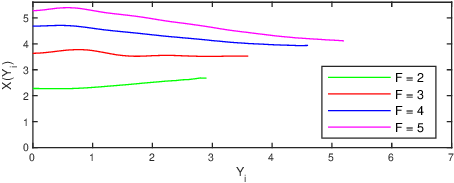}\\[-0.5em]
\caption{Discard policies learned by Algorithm~\ref{alg:2} for the sum of the AoI penalty cost 
(defined by the identity function) and a transmission cost, for several values of $F$.}\vspace{-10pt}
\label{fig:polcost}
\end{center}
\end{figure}

Figure~\ref{fig:polcost} shows the discard policies learned by the  Algorithm~\ref{alg:2} 
for several values of $F$. 
%As expected, the higher the transmission cost $F$, %the higher are the values of maximum delays $X_i$ of the policy.
As expected, increasing the transmission cost parameter $F$ shifts the optimal policy toward selecting larger maximum delays $X_i$.
This is because the learned policy aims to balance  the AoI cost,
which can be reduced by decreasing  $X_i$,
with the transmissions cost,
which can be minimized  by increasing $X_i$, consequently reducing the number of transmissions.

\subsection{Combined use of wait and discard strategies}\label{sec:twostrategies}

\noindent This subsection illustrates the joint application of the wait and discard strategies through Algorithm~\ref{alg:3}.

\begin{figure}[tbp]
\centering
\subfloat[]{\includegraphics[width=\columnwidth]{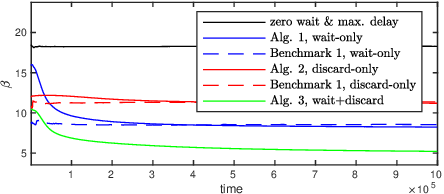}}\\\subfloat[]{\includegraphics[width=\columnwidth]{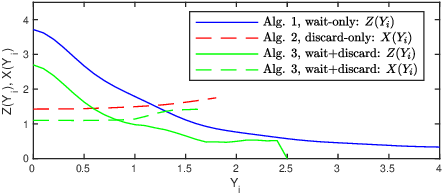}}
\caption{(a) Time-average costs with several strategies: wait, discard and both simultaneously. (b) Policies.
% when strategies used alone and when used simultaneously.
\label{fig:cost3waitdiscard}}
\end{figure}

Figure~\ref{fig:cost3waitdiscard} shows an example of using simultaneously the wait strategy and the discard strategy, 
for the sum of the AoI penalty cost (defined by the identity function) and a transmission cost,
\begin{equation}
C_i=k_iF+\int_{Y_{i-1}}^{Y_{i-1}+Z_i+(k_i-1)X_i+Y_i}t^{\gamma}\; dt,
\end{equation}
with $\gamma=1.5$, $F=4$, and the log-normal channel model with $\rho=0.5$.
In Figure~\ref{fig:cost3waitdiscard}\,(a),
 the time-average cost obtained 
by Algorithm~\ref{alg:3} applying both strategies together 
is compared with that obtained by Algorithm~\ref{alg:1} and Benchmark~1
when applying the wait only strategy, and
with that obtained by Algorithm~\ref{alg:2} and Benchmark~1
when applying the discard only strategy.
Figure~\ref{fig:cost3waitdiscard}\,(b) shows the policy for the three setups described above: \textit{wait-only} policy for Algorithm~\ref{alg:1} (blue curve), \textit{discard-only} polices for Algorithm~\ref{alg:2} (red curve), and \textit{wait+discard} policies for Algorithm~\ref{alg:3} (green curves).
In the \textit{wait+discard} setup, both strategies are applied simultaneously: the solid green curve represents the configuration of the wait policy, while the dashed green curve shows the configuration of the discard policy.
As can be seen, the use combined of the wait and discard strategies outperforms the only wait and the only discard strategies.

\subsection{Simplified channel model}\label{sec:simplified_channel_model}

\noindent In this section we compare the
performance of Algorithm~\ref{alg:1} and Algorithm~\ref{alg:2} to the optimal policy in scenarios different from the particular case of~\eqref{eq:costPenaltyF0}, where it is impossible to calculate the optimal policy by means of the method proposed in~\cite{sun17}.
Therefore, to obtain the optimal policy numerically,
we define a simplified channel
model where the transmission delay is defined by a Gilbert-Elliot model, which is a two-state
discrete-time Markov chain widely used for modelling error patterns with temporal
correlation in transmission channels~\cite{hass08}.
Let $\{0,1\}$ be the set of states, and let the transition matrix be
\begin{equation}\nonumber
\begin{bmatrix}
1-p & p  \\
q   & 1-q
\end{bmatrix}.
\end{equation}
In this model, state $0$ represents a {\em good} state, while state $1$ represents a {\em bad} state for the communication system, including the physical channel and the mechanisms implemented at the different layers of the protocol stack.

For simplicity, let us assume that at each state the delays are fixed: 
$Y_i=y_0$ in state $0$ and $Y_i=y_1$ in state $1$, such that $y_0\in [0,y_1)$.
Note that both the set of the system states $\{y_0,y_1\}$ and the set of actions $\{Z(y_0),Z(y_i)\}$ are discrete and finite with this channel model,
while  Algorithm~\ref{alg:1} and Algorithm~\ref{alg:2} aim to learn a continuous policy in a continuous state space.
When run with this channel model, these algorithms will only learn the policy for the two states $\{y_0,y_1\}$, 
since these are the only states that are visited.
For each state, the learned policy will consist of a random variable
whose mean should approach the corresponding value in the optimal policy.

\begin{figure}[tbp]
\begin{center}
\includegraphics[width=\columnwidth]{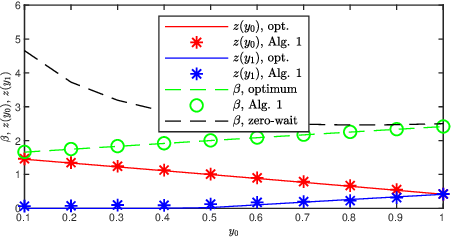}
\caption{Time-average costs and optimal policies for the wait strategy in the two-state Markov chain delay model.
	%Time-average costs and wait strategies in the two states Markov chain delay model.
	}
\label{fig:valwaitcostid}
\end{center}
\end{figure}

In the first experiment, the wait strategy is used and 
the cost function is given by 
the sum of the AoI penalty cost (defined by the identity function $p(\Delta(t))=\Delta(t)$) and
a transmission cost~\eqref{eq:idcostwait}. 
For this case, we have solved the finite Markov Decision Problem to obtain a closed expression for the average cost, to subsequently numerically obtain the values of $Z(y_0)$ and $Z(y_1)$ that minimize this function.
Figure~\ref{fig:valwaitcostid} shows the comparison of the results calculated numerically with the results obtained with Algorithm~\ref{alg:1} for
$F=1$, $p=0.01$, $q=0.04$, $y_1=1$ and several values of $y_0$ in the range $0.1\leq y_0\leq 1$.
The solid lines represent the values of the optimal policy 
and the asterisks are the respective values obtained by Algorithm~\ref{alg:1}. 
The dashed lines are the time-average costs obtained with the optimal wait policy and
with the zero-wait policy, and the circles represent the time-average costs computed by Algorithm~\ref{alg:1}. 
It can be seen in Figure~\ref{fig:valwaitcostid} that the policy values for both states obtained by Algorithm~\ref{alg:1} are very close to the optimum.

\begin{figure}[tbp]
\begin{center}
\includegraphics[width=\columnwidth]{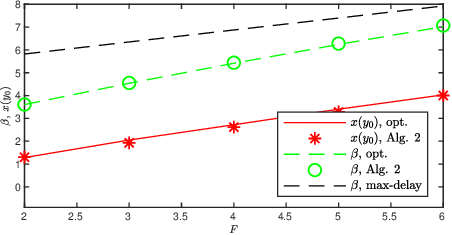}
\caption{Time-average costs and optimal policies for the discard strategy in the two-state Markov chain delay model.}
\label{fig:valdiscostid}
\end{center}
\end{figure}

The second experiment has been conducted with the discard policy and the cost function given by 
the sum of the AoI penalty cost (defined by the identity function) and
a transmission cost~\eqref{eq:idcostdiscF}. 
Figure~\ref{fig:valdiscostid} shows the comparison of the results calculated numerically with the results obtained with Algorithm~\ref{alg:2} for
$p=0.1$, $q=0.9$, $y_0=1$,  $y_1=10$ and several values of $F$ in the range $2\leq F\leq 7$.
With this strategy and channel model, for each value of $F$ 
the optimal policy for state $y_0$ is a value $X^{\text{opt}}_0,\; y_0<X^{\text{opt}}_0<y_1$.
Since $X^{\text{opt}}_0<y_1$, 
the last transmission delay at every delivery is always $y_0$
and therefore 
the optimal policy for $y_1$ is meaningless.
The value of $X^{\text{opt}}_0$ for each value of $F$ has been calculated numerically 
and plotted as a solid line.
The asterisks are the respective values obtained by Algorithm~\ref{alg:2},
the dashed lines are the time-average costs obtained with the optimal policy
and with the `maximum delay policy', 
and the circles represent the time-average costs computed by Algorithm~\ref{alg:2}. 
It can be seen that the policy values obtained by Algorithm~\ref{alg:2} are very close to the  optimal ones,
 although due to the random nature of the learned policy, 
 they present a small deviation that does not have an appreciable impact on the resulting time-average cost.

\subsection{Generalization for a two-way delay channel model}\label{sec:generalization}

\begin{figure}[tbp]
	\begin{center}
		\resizebox{0.8\columnwidth}{!}{
			\begin{tikzpicture}
				\draw[very thick,-To](0.9,0)--(2.6,0);
				\draw[very thick,-To](6,-0.9)--(4.9,-0.9);
				\draw[very thick,-To](2.4,-0.9)--(0.9,-0.9);
				\draw[very thick,-To](4.8,0)--(6,0);
				\node [above,very thick,above,rectangle,inner sep=2pt,draw,anchor=center,align=center, minimum height=10mm ] at (3.7,0.1) {forward delay};
				\node [above,very thick,above,rectangle,inner sep=2pt,draw,anchor=center,align=center, minimum height=10mm ] at (3.7,-1) {backward delay};
				\node [very thick,above,circle,,inner sep=8pt,draw,anchor=center] at (0.1,-0.4) {source};
				\node [very thick,above,circle,inner sep=1pt,draw,anchor=center] at (6.8,-0.4) {destination};
				\node [anchor=center]at (1.6,0.3){data};
				\node [anchor=center]at (1.7,-0.6){feedback};
		\end{tikzpicture}}
		\caption{Two-way delay data delivery model.}\label{fig:modelF}
	\end{center}
\end{figure}
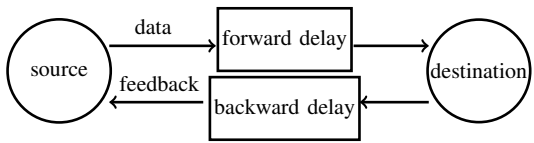

\noindent
The generalization of the proposed algorithm to a two-way delay channel model is straightforward. 
Consider the data delivery model shown in Fig.~\ref{fig:modelF}, 
which includes a delay between the delivery of each data unit and the reception of the corresponding feedback message. 
Let $Y'_i$ denote the feedback delay associated with the $i$-th delivery, 
measured from the time the data unit is delivered at the destination until the feedback is received at the source.

With this model, the width of the interval $(D_{i-1}, D_i]$ in the wait strategy becomes $D_i - D_{i-1} = Y'_{i-1} + Z_i + Y_i$, 
and in the discard strategy, $D_i - D_{i-1} = Y'_{i-1} + (k_i - 1)X_i + Y_i$. 
To accommodate this model, the policy gradient algorithm is updated by redefining the system state as
\begin{equation}
	S_i \triangleq (Y_{i-1}, Y'_{i-1}), \quad Y_{i-1} \in [0, \infty), \quad Y'_{i-1} \in [0, \infty).
\end{equation}
With this two-dimensional state space, the linear approximation of the mean of the policy distribution in~\eqref{eq:fourier} becomes
\begin{equation}\label{eq:fourier2}
	\mu(Y, Y', \vec{\theta}) = \sum_{j=0}^{d_1-1} \sum_{k=0}^{d_2-1} \theta^{jk} \cos\left(\frac{j\pi Y}{Y_{\text{max}}} + \frac{k\pi Y'}{Y'_{\text{max}}}\right),
\end{equation}
where $\vec{\theta}$ contains $d_1 \times d_2$ elements. Similarly, the state-value function $\hat{\nu}(Y, \vec{\omega})$ is approximated by a two-dimensional Fourier series.

\textit{Benchmark~2}, presented in\cite{tsai23}, can also operate under this model, but only when the wait strategy is used and the cost function follows~\eqref{eq:costPenaltyF0}.

\begin{figure}[tbp]
	\begin{center}
		\includegraphics[width=\columnwidth]{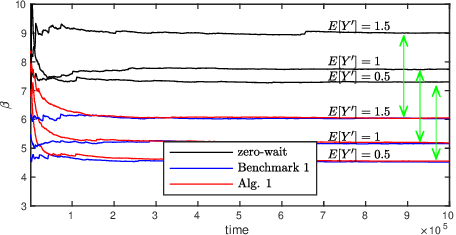}
		\caption{Time-average costs for wait strategy and AoI penalty cost with identity function, for backward delays with means 0.5, 1 and 1.5 respectively.}
		\label{fig:generalization}
	\end{center}
\end{figure}

Figure~\ref{fig:generalization} shows the results for Algorithm~\ref{alg:1} and \textit{Benchmark~2} under the wait strategy with the AoI penalty cost defined by the identity function. The backward delay $Y'$ is modeled as a log-normally distributed Markov process with three different mean values: $0.5$, $1$, and $1.5$. The mean of $Y$ is fixed at $1$ in all cases.

As in the one-way delay model, the performance of Algorithm~\ref{alg:1} closely matches that of \textit{Benchmark~2}, and it clearly outperforms the \textit{zero-wait} policy.

\subsection{Computational cost}\label{sec:comp_cost}

\noindent A comparison of the computational cost of the algorithms used in Section~\ref{sec:results}
is shown in Table~\ref{tab:compcost}.
For each algorithm, the time spent to process $10^6$ steps has been measured and normalized by the execution time of Algorithm~\ref{alg:1}, which is the one with the lowest computational cost. 
All algorithms have been tested on the same platform, with the same API and sharing most of the software modules. 
Although the exact processing time depends on the choice of the algorithm parameters and other factors,
the times of the different algorithms differ from each other significantly  
and this provides insight into their relative complexity.
Algorithms~\ref{alg:1} and~\ref{alg:2} are clearly faster than the others because they only estimate and utilize one-dimensional functions: 
one policy in Algorithm~\ref{alg:1}, 
one policy and one state-value function in Algorithm~\ref{alg:2}, 
and two policies along with one state-value function when implementing two strategies simultaneously. 
In contrast, Benchmark~1 requires more than an order of magnitude more time because it estimates a two-dimensional action-value function and, at each step, performs a search in this function to determine the optimal action. Benchmark~2, on the other hand, must process an online register of $Y_i$~samples. 
If the penalty function is known, it takes about an order of magnitude longer. 
However, if the penalty function is unknown and needs to be estimated online, 
its complexity increases further, reaching up to three orders of magnitude higher.

\begin{table}
\centering
\caption{Normalized execution time of the tested algorithms.}\label{tab:compcost}\vspace{-5pt}
\begin{footnotesize}
\begin{tabularx}{\columnwidth}{Xr}
\multicolumn{2}{r}{\textbf{Normalized execution time}} \\ \hline
Algorithm 1 &	1  \\
Algorithm 2 &		3   \\
Algorithm 1/2  (2 strategies simultaneously) &		4   \\
Benchmark~1	&	43   \\
Benchmark~2	(penalty function known) &	18   \\
Benchmark~2 (penalty function estimated) & $2132$   \\\hline
\end{tabularx}
\end{footnotesize}
\end{table}

\section{Conclusions}\label{sec:conclusion}

\noindent We present two versions of a policy gradient algorithm to reduce the time-average AoI cost,
that can be run online 
and does not require any knowledge of the statistics of the transmission delays,
nor of the characteristics of the function that quantifies the cost of data deliveries.
This algorithm  accepts a quite general definition  of the cost function,
even when this function cannot be characterized as the integral of any penalty function,
and also when it includes a transmission cost.
Also, it works for different strategies to reduce the time-average cost.
It has been tested for several cost functions,
and for two distinct update strategies to reduce time-average cost:
a wait strategy and a discard strategy.
The algorithm has been  validated by comparing its results with the theoretical optimum when available, 
showing that the resulting time-average cost reduction is very close to the optimal.
For more general definitions of cost and strategies for
which the theoretical optimum is not available, 
the algorithm performance has been compared with that of a state-of-the-art algorithm based on SARSA, 
finding that it obtains a significantly greater time-average cost reduction, 
and has a computational cost that is at least one order of magnitude lower.
Furthermore, the algorithm can be used to implement several strategies simultaneously
with only a linear increase of the computational cost, 
and, with minor modifications, it may potentially be valid for strategies other than the ones described here.

\section*{Acknowledgments}

This work was supported through Grant PID2021-123168NB-I00, funded by MCIN/AEI/10.13039/50 1100011033 and the European Union A way of making Europe/ERDF, and Grant TED2021-131387B-I00, funded by MCIN/AEI/10.13039/501100011033 and the European Union NextGenerationEU/RTRP. 
%-------------------------------------------------------------------------------------------------
%-------------------------------------------------------------------------------------------------
%\bibliographystyle{IEEEtran}
%\bibliography{bib}

%
% Generated by IEEEtran.bst, version: 1.14 (2015/08/26)

\begin{IEEEbiography}[{\includegraphics[width=1in,height=1.25in,clip,keepaspectratio]{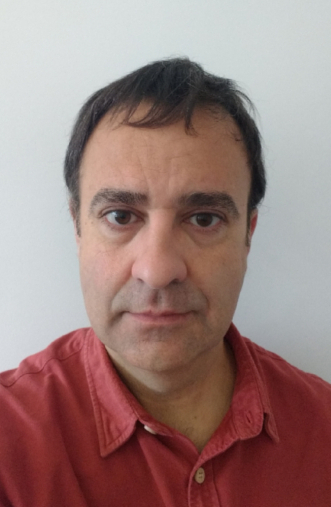}}]{Jos\'e-Ram\'on~Vidal}
received the Telecommunication Engineering Ph.D. degree from the Universitat Polit\`ecnica de Val\`encia(UPV), Spain. He is currently an Associate Professor at the Department of Communicactions, Higher Technical School of Telecommunication Engineering, UPV. His current research interest includes the application of game theory and machine learning to resource allocation in cognitive radio networks and to economic modeling of telecommunication service provision. In these areas, he has authored or coauthored several papers in refereed journals and conference proceedings.
\end{IEEEbiography}

\begin{IEEEbiography}[{\includegraphics[width=1in,height=1.25in,clip,keepaspectratio]{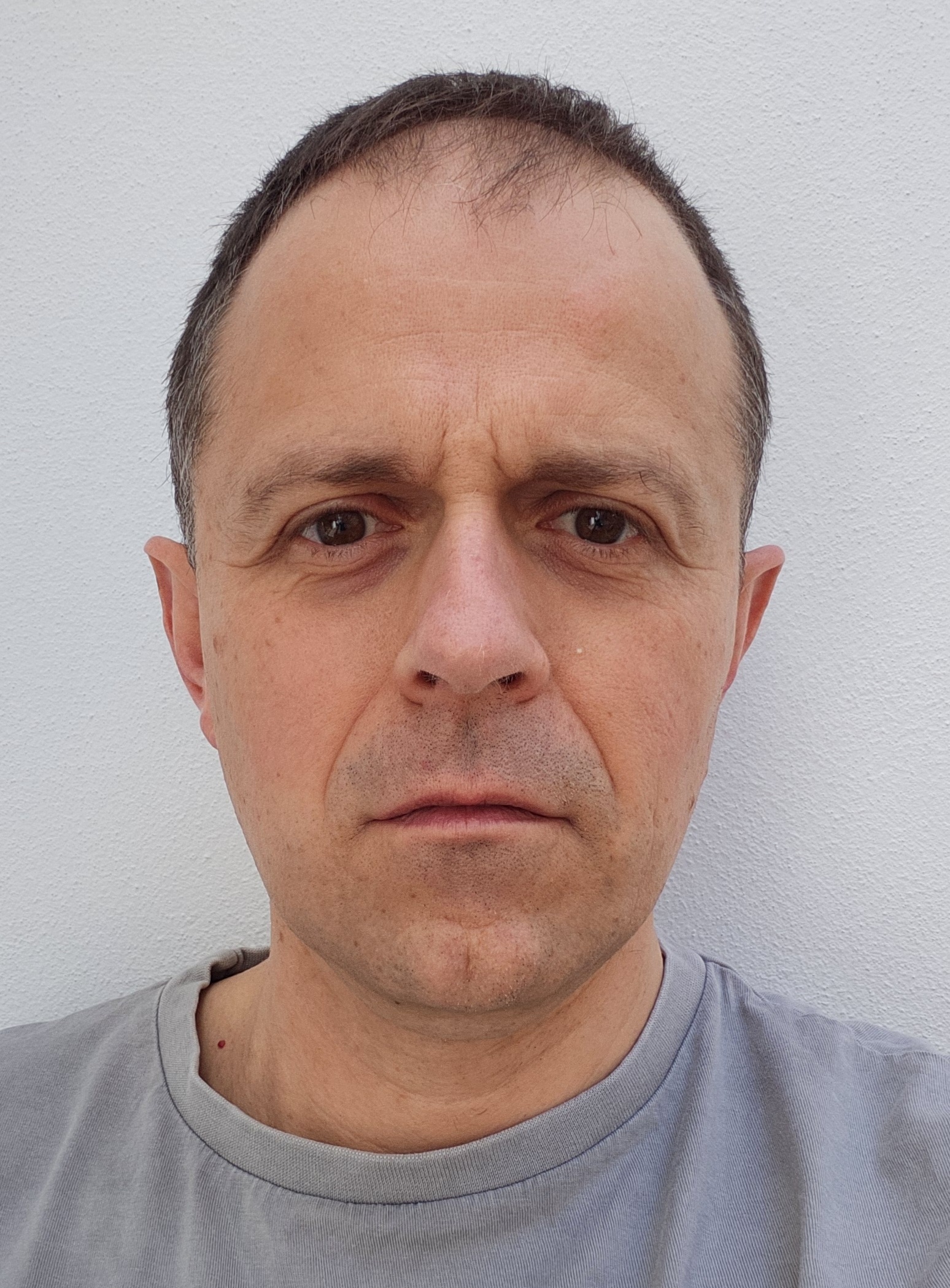}}]{Vicent Pla}
received the degree in Telecommunication Engineering (BE+ME equivalent) and the PhD degree in Telecommunications from the Universitat Polit\`ecnica de Val\`encia, Spain, in 1997 and 2005, respectively, and the BSc degree in Mathematics from the Universidad Nacional de Educaci\'on a Distancia, Spain, in 2015. In 1999, he joined the Department of Communications at the Universitat Polit\`ecnica de Val\`encia, where he is currently a Professor. His research interests focus on the modeling and performance analysis of communication networks, with an emphasis on traffic and resource management in wireless systems. In these areas, he has authored numerous papers in refereed journals and conference proceedings, and has participated in or led multiple research projects. He has conducted research visits at the University of Pennsylvania, USA; Aalto University, Finland; the University of Manitoba, Canada; Huazhong University of Science and Technology, China; and Aalborg University, Denmark.

\end{IEEEbiography}
\begin{IEEEbiography}[{\includegraphics[width=1in,height=1.25in,clip,keepaspectratio]{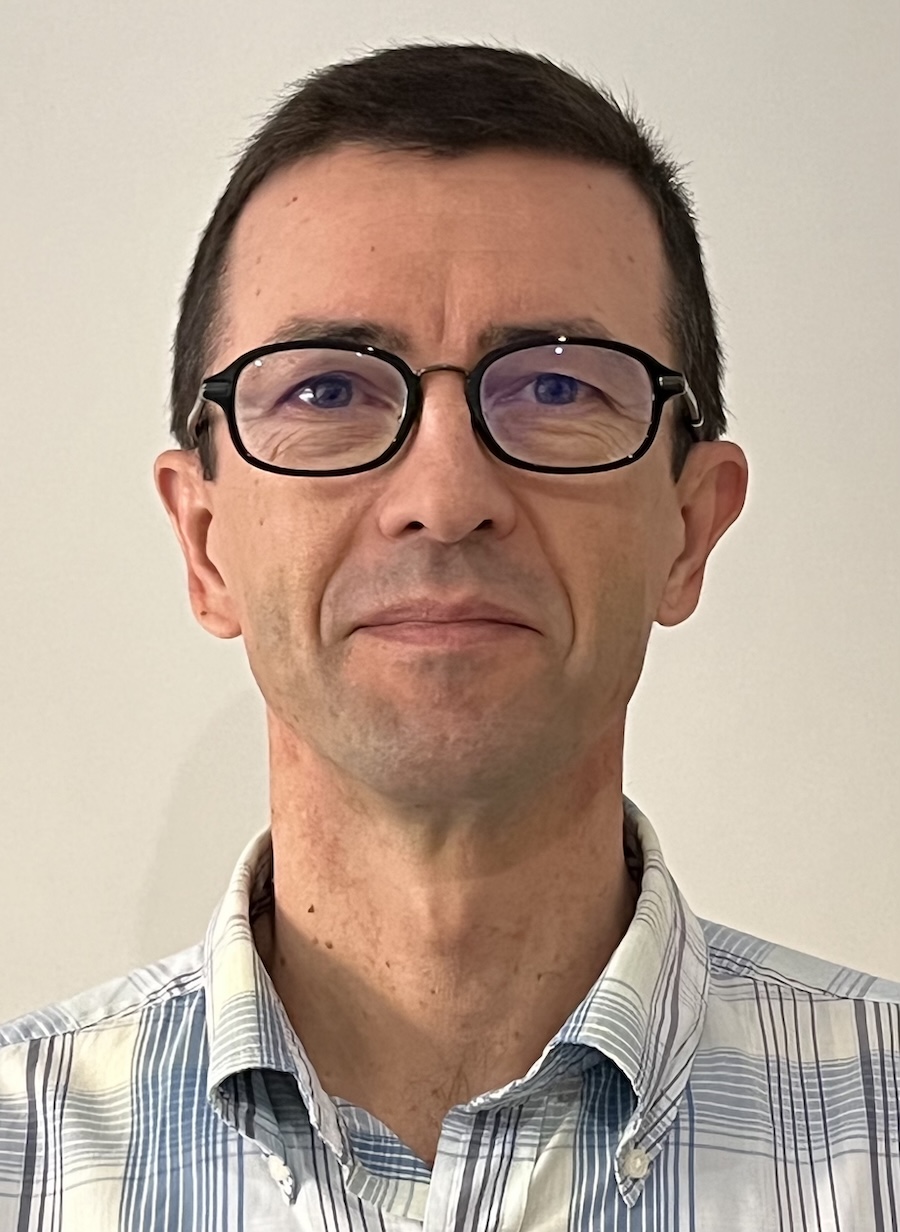}}]{Luis Guijarro}
received the M.Eng. and Ph.D. degrees in telecommunications from the Universitat Polit\`ecnica de Val\`encia (UPV), Spain, in 1993 and 1998. He is a Professor in telecommunications economics and regulation with the Department of Communications, UPV. He has coauthored the book Electronic Communications Policy of the European Union (UPV, 2010). He has researched in traffic management in ATM networks and in e-Government and his current research interests include economic modeling of telecommunication service provision. He has published in refereed journals and conferences proceedings in the topics of peer-to-peer interconnection, cognitive radio networks, net neutrality, wireless sensor networks, 5G, 6G, and platform economics.
\end{IEEEbiography}
 
\begin{IEEEbiography}[{\includegraphics[width=1in,height=1.25in,clip,keepaspectratio]{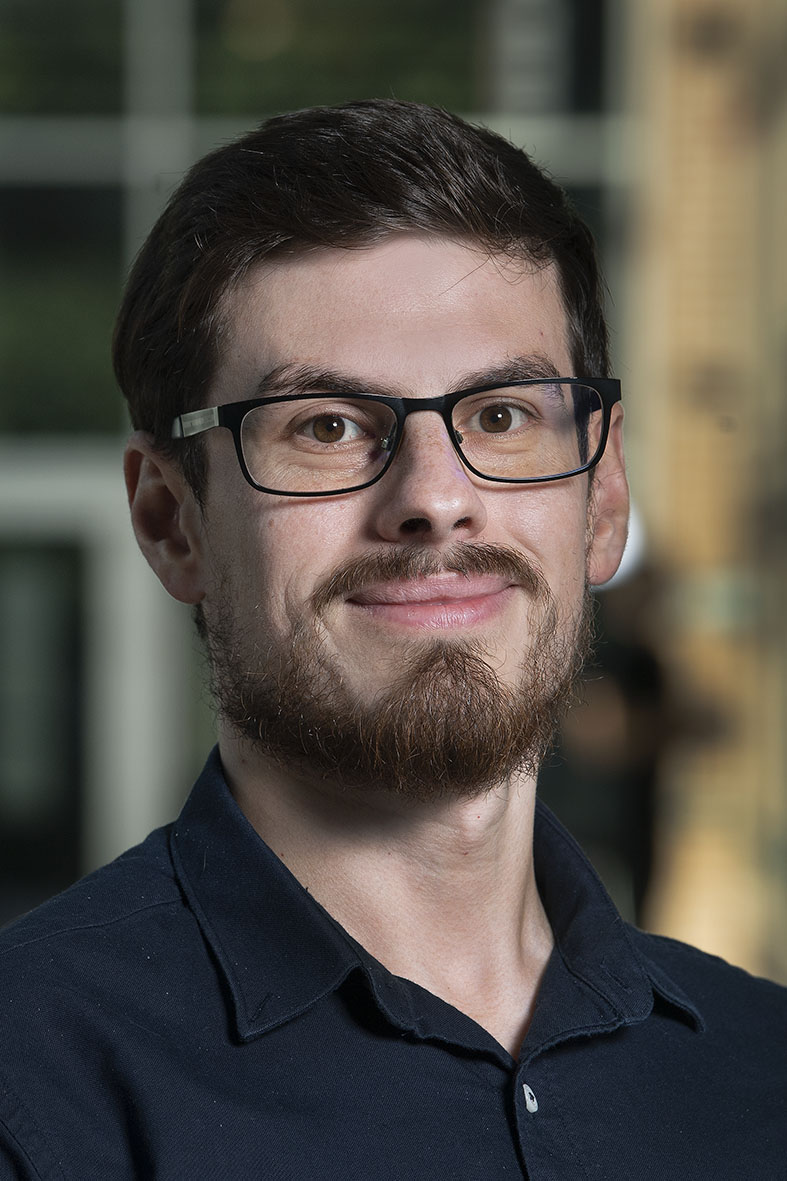}}]{Israel Leyva-Mayorga}
(Member, IEEE) received the B.Sc. degree in telematics engineering and the M.Sc. degree (Hons.) in mobile computing systems from the Instituto Polit\'ecnico Nacional (IPN), Mexico, in 2012 and 2014, respectively, and the Ph.D. degree (cum laude and extraordinary prize) in telecommunications from the Universitat Polit\`ecnica de Val\`encia (UPV), Spain, in 2018. He was a Visiting Researcher at the Department of Communications, UPV, in 2014, and at the Deutsche Telekom Chair of Communication Networks, Technische Universit\"at Dresden, Germany, in 2018. He is currently an Associate Professor at the Connectivity Section (CNT) of the Department of Electronic Systems, Aalborg University (AAU), Denmark, where he served as a Postdoctoral Researcher from January 2019 to July 2021. He is an Associate Editor for IEEE WIRELESS COMMUNICATIONS LETTERS, where he received the Best Editor award in 2025, a recipient of the Novo Nordisk Foundation Prize for Excellence in Technical Science Teaching 2025, and a Board Member for one6G. His research interests include beyond-5G and 6G networks, satellite communications, and random and multiple access protocols.
\end{IEEEbiography}

\end{document}